\documentclass[fleqn,usenatbib]{mnras}

\usepackage[T1]{fontenc}

\DeclareRobustCommand{\VAN}[3]{#2}
\let\VANthebibliography\thebibliography
\def\thebibliography{\DeclareRobustCommand{\VAN}[3]{##3}\VANthebibliography}

\usepackage{graphicx}	% Including figure files
\usepackage{amsmath}	% Advanced maths commands
\usepackage{amssymb}	% Extra maths symbols
\usepackage{newtxtext,newtxmath}

\usepackage{lineno}
\usepackage{natbib}

\usepackage{nameref}
\usepackage{orcidlink}
\title[Neutrino follow-up with ZTF]{Neutrino follow-up with the Zwicky Transient Facility: Results from the first 24 campaigns}

\author[R. Stein et al.]{Robert Stein \orcidlink{0000-0003-2434-0387}$^{1, 2, 3}$\thanks{E-mail: rdstein@caltech.edu},
Simeon Reusch \orcidlink{0000-0002-7788-628X}$^{1, 2}$,
Anna Franckowiak \orcidlink{0000-0002-5605-2219}$^{1, 2, 4}$,
Marek Kowalski \orcidlink{0000-0001-8594-8666}$^{1, 2}$,
\newauthor
Jannis Necker \orcidlink{0000-0003-0280-7484}$^{1, 2}$,
Sven Weimann $^{4}$,
Mansi M. Kasliwal \orcidlink{0000-0002-5619-4938}$^{3}$,
Jesper Sollerman \orcidlink{0000-0003-1546-6615}$^{5}$,
Tomas Ahumada \orcidlink{0000-0002-2184-6430}$^{6,7}$\thanks{LSSTC Data Science Fellow},
\newauthor
Pau Amaro Seoane \orcidlink{0000-0003-3993-3249}$^{8,9,10}$,
Shreya Anand \orcidlink{0000-0003-3768-7515}$^{3}$,
Igor Andreoni \orcidlink{0000-0002-8977-1498} $^{11, 12, 13}$,
Eric C. Bellm \orcidlink{0000-0001-8018-5348}$^{14}$,
\newauthor
Joshua S. Bloom \orcidlink{0000-0002-7777-216X}$^{15, 16}$,
Michael Coughlin \orcidlink{0000-0002-8262-2924}$^{17}$,
Kishalay De $^{18}$,
Christoffer Fremling \orcidlink{0000-0002-4223-103X}$^{3}$,
Suvi Gezari \orcidlink{0000-0003-3703-5154}$^{19}$,
\newauthor
Matthew Graham \orcidlink{0000-0002-3168-0139}$^{3}$,
Steven L. Groom \orcidlink{0000-0001-5668-3507}$^{20}$,
George Helou \orcidlink{0000-0003-3367-3415}$^{20}$,
David L. Kaplan \orcidlink{0000-0001-6295-2881}$^{21}$,
Viraj Karambelkar \orcidlink{0000-0003-2758-159X}$^{3}$,
\newauthor
Albert K.H. Kong \orcidlink{0000-0002-5105-344X}$^{22}$,
Erik C. Kool \orcidlink{0000-0002-7252-3877}$^{5}$,
Massimiliano Lincetto \orcidlink{0000-0002-1460-3369}$^{4}$,
Ashish~A.~Mahabal \orcidlink{0000-0003-2242-0244}$^{3, 23}$,
\newauthor
Frank J. Masci \orcidlink{0000-0002-8532-9395}$^{20}$,
Michael S. Medford \orcidlink{0000-0002-7226-0659}$^{15,16}$,
Robert Morgan \orcidlink{0000-0002-7016-5471}$^{24}$,
Jakob Nordin \orcidlink{0000-0001-8342-6274}$^{2}$,
Hector Rodriguez $^{25}$,
\newauthor
Yashvi Sharma \orcidlink{0000-0003-4531-1745}$^{3}$,
Jakob van Santen \orcidlink{0000-0002-2412-9728}$^{1}$,
Sjoert van Velzen \orcidlink{0000-0002-3859-8074}$^{26}$,
Lin Yan \orcidlink{0000-0003-1710-9339}$^{25}$
\\
$^{1}$Deutsches Elektronen-Synchrotron (DESY), Platanenallee 6, D-15378 Zeuthen, Germany\\
$^{2}$Institut fur Physik, Humboldt-Universit\"at zu Berlin, D-12489 Berlin, Germany\\
$^{3}$Division of Physics, Mathematics, and Astronomy, California Institute of Technology, Pasadena, CA 91125, USA\\
$^{4}$Fakult\"at f\"ur Physik \& Astronomie, Ruhr-Universit\"at Bochum, D-44780 Bochum, Germany\\
$^{5}$Oskar Klein Centre, Department of Astronomy, Stockholm University, 10691 Stockholm, Sweden\\
$^{6}$Department of Astronomy, University of Maryland, College Park, MD 20742, USA\\
$^{7}$Astrophysics Science Division, NASA Goddard Space Flight Center, Mail Code 661, Greenbelt, MD 20771, USA\\
$^{8}$Max Planck Institute for Extraterrestrial Physics, Garching, Germany\\
$^{9}$Institute of Applied Mathematics, Academy of Mathematics and Systems Science, CAS, Beijing, China\\
$^{10}$Kavli Institute for Astronomy and Astrophysics, Beijing, China\\
$^{11}$Joint Space-Science Institute, University of Maryland, College Park, MD 20742, USA\\
$^{12}$Department of Astronomy, University of Maryland, College Park, MD 20742, USA\\
$^{13}$Astrophysics Science Division, NASA Goddard Space Flight Center, Mail Code 661, Greenbelt, MD 20771, USA\\
$^{14}$DIRAC Institute, Department of Astronomy, University of Washington, 3910 15th Avenue NE, Seattle, WA 98195, USA\\
$^{15}$Department of Astronomy, University of California, Berkeley, CA 94720-3411, USA\\
$^{16}$Lawrence Berkeley National Laboratory, 1 Cyclotron Road, MS 50B-4206, Berkeley, CA 94720, USA\\
$^{17}$School of Physics and Astronomy, University of Minnesota, Minneapolis, MN 55455, USA\\
$^{18}$MIT-Kavli Institute for Astrophysics and Space Research, 77 Massachusetts Ave., Cambridge, MA 02139, USA\\
$^{19}$Space Telescope Science Institute, 3700 San Martin Dr., Baltimore, MD 21218, USA\\
$^{20}$IPAC, California Institute of Technology, 1200 E. California, CA 91125, USA\\
$^{21}$Center for Gravitation, Cosmology and Astrophysics, Department of Physics, University of Wisconsin-Milwaukee, Milwaukee, WI 53201\\
$^{22}$Institute of Astronomy, National Tsing Hua University, Hsinchu 30013, Taiwan\\
$^{23}$Center for Data Driven Discovery, California Institute of Technology, Pasadena, CA 91125, USA\\
$^{24}$Physics Department, University of Wisconsin-Madison, Madison, WI 53706, USA\\
$^{25}$Caltech Optical Observatories, California Institute of Technology, Pasadena, CA 91125, USA\\
$^{26}$Leiden Observatory, Leiden University, Postbus 9513, 2300 RA, Leiden, The Netherlands\\
}

\date{Accepted 05 March 2023. Received 25 February 2023; in original form 08 April 2022}

\pubyear{2023}

\begin{document}
\label{firstpage}
\pagerange{\pageref{firstpage}--\pageref{lastpage}}
\maketitle
\clearpage
% Abstract of the paper
\begin{abstract}

The Zwicky Transient Facility (ZTF) performs a systematic neutrino follow-up program, searching for optical counterparts to high-energy neutrinos with dedicated Target-of-Opportunity (ToO) observations. Since first light in March 2018, ZTF has taken prompt observations for 24 high-quality neutrino alerts from the IceCube Neutrino Observatory, with a median latency of 12.2 hours from initial neutrino detection. From two of these campaigns, we have already reported tidal disruption event (TDE) AT 2019dsg and likely TDE AT 2019fdr as probable counterparts, suggesting that TDEs contribute >7.8\% of the astrophysical neutrino flux. We here present the full results of our program through to December 2021. No additional candidate neutrino sources were identified by our program, allowing us to place the first constraints on the underlying optical luminosity function of astrophysical neutrino sources. Transients  with optical absolutes magnitudes brighter that $-21$ can contribute no more than 87\% of the total, while transients brighter than $-22$ can contribute no more than 58\% of the total, neglecting the effect of extinction and assuming they follow the star formation rate. These are the first observational constraints on the neutrino emission of bright populations such as superluminous supernovae. None of the neutrinos were coincident with bright optical AGN flares comparable to that observed for TXS 0506+056/IC170922A, with such optical blazar flares producing no more than 26\% of the total neutrino flux. We highlight the outlook for electromagnetic neutrino follow-up programs, including the expected potential for the Rubin Observatory.
\end{abstract}

% Select between one and six entries from the list of approved keywords.
% Don't make up new ones.

%%%%%%%%%%%%%%%%%%%%%%%%%%%%%%%%%%%%%%%%%%%%%%%%%%

%%%%%%%%%%%%%%%%% BODY OF PAPER %%%%%%%%%%%%%%%%%%

\section{Introduction}

Astrophysical neutrinos are produced through the interaction of accelerated hadrons with matter or photons. A flux of  astrophysical neutrinos with energies in the TeV-PeV range, was first discovered by IceCube in 2013 \citep{icecube_discovery_13}. Recent results suggest that a substantial fraction of these high-energy neutrinos are produced in the cores of Active Galactic Nuclei (AGN) \citep{ic_agn_21}, with additional evidence for neutrino emission from the nearby AGN NGC 1068 \citep{ic_ps_10_yr}. Beyond this static component, various transient or variable source classes have been proposed as possible contributors to the neutrino flux, including gamma-ray bursts (GRBs) \citep{waxman_bahcall_97_grb}, core-collapse supernovae (CCSNe) \citep{murase_csm_sn_11}, TDEs \citep{farrar_09} and blazars \citep{mannheim_93}.  All of these proposed neutrino source classes have electromagnetic signatures at optical wavelengths.

To aid in identifying these time-varying source candidates, \hbox{IceCube} has operated an automated program since 2016 to publish realtime high-energy neutrino alerts \citep{ic_realtime_17}, enabling contemporaneous electromagnetic observations of putative neutrino source candidates at radio \citep{telamon_21}, optical \citep{kowalski_07, ptf_15, ps_icecube_19, decam_ic_19, necker_22}, X-ray \citep{swift_15, integral_21}, and gamma-ray wavelengths \citep{agile_ic_19, garrappa_21, satalecka_21}. In 2017, this realtime program led to the identification of a flaring blazar, TXS 0506+056, as the likely source of high-energy neutrino IC170922A \citep{ic_txs_mm_18}. Studies of these high-energy neutrino alerts have suggested possible correlations with blazar sub-populations, namely radio-bright blazars \citep{plavin_20, plavin_21} and intermediate-energy/high-energy peaked blazars (IBLs/HBLs) \citep{giommi_20b}.

The Zwicky Transient Facility (ZTF) is an optical telescope with a 47 sq. deg field of view \citep{ztf_system, ztf_obs_system}. Since first light in 2018, ZTF has operated a dedicated neutrino follow-up program, in which the arrival directions of IceCube neutrino alerts are observed with Target-of-Opportunity (ToO) observations \citep{ztf_19_science}. This program has led to the identification of two further likely high-energy neutrino sources, the TDE AT 2019dsg \citep{bran} and the probable TDE AT 2019fdr (\citealt{tywin}, though see \citealt{pitik_21} for an alternative interpretation). Accounting for the contribution of higher-redshift sources, these results suggest that at least 7.8\% of neutrino alerts arise from the broader TDE population \citep{tywin}. Archival analysis of ZTF data revealed further evidence of a correlation between such flares and high-energy neutrinos \citep{van_velzen_21}. 

In this paper we outline the full results of the ZTF neutrino follow-up program, which has to date included 24 dedicated neutrino follow-up campaigns. This sample enables novel constraints to be set on the neutrino emission of a broad range of optical transient and variable populations. 

The paper is organised as follows: Section \ref{sec:selection} outlines the program itself, including trigger criteria and optical candidate selection. Section \ref{sec:transients} outlines transient candidates identified by the program, and subsequent electromagnetic observations to determine their nature. Section \ref{sec:AGN flares} outlines optical AGN flares found coincident with neutrinos, and Section \ref{sec:literature} provides data on two candidate neutrino sources identified in the literature. Section \ref{sec:limits} considers the various constraints that can be placed on different possible neutrino source populations from our program. Section \ref{sec:conclusion} summarises the main results, and outlines how such follow-up programs may improve with future observatories.

%--------------------------------------------------------------------
\section{Neutrino Follow-up with ZTF}
\label{sec:selection}

Neutrino alerts are generally published by IceCube in the form of automated Gamma-ray Coordination Network (GCN) Notices\footnote{\url{https://gcn.gsfc.nasa.gov}}, with initial estimates of the statistical uncertainty on the neutrino position. These positions are then superseded after a few hours by a GCN Circular with an updated localisation that also incorporates systematic uncertainties \citep{cristina_icrc_21}. Given the substantial increase in localisation area once systematic effects are accounted for, with increases of factor 5 not being uncommon, we rely on the latter category to perform our search for neutrino counterparts. 

With ZTF, we aim to observe all accessible high-quality neutrino alerts from IceCube. We define high-quality alerts as those with a high probability to be of astrophysical origin (`signalness' > 50\%), or those which are well-localised (a 90\% localisation area $<$ 10 sq.\,deg.).  Though IceCube labels alerts as Gold or Bronze based on average quality, individual Bronze alerts have been reported with signalness values greater than 50\% (e.g. IC211208A) and Gold alerts have been reported with signalness values less than 15\% (e.g. IC201130A). We therefore ignore the labelling of these streams, and select exclusively based on the signalness and localisation.

We have followed up 24 neutrinos in the period from survey start on 2018 March 20 to 2021 December 31, out of a total of 79 neutrino alerts published by IceCube during that time. Table \ref{tab:nu_alerts} summarises each neutrino alert  observed by ZTF. From 2019 June 17, IceCube published neutrino alerts with improved selection criteria (V2) to provide an elevated alert rate \citep{ic_realtime_19}. In addition to 1 of the 12 alerts under the old selection, ZTF followed up 23 of the 67 alerts published under the V2 selection. Midway through the ZTF program, an additional cut on neutrino alert galactic latitude (|b| > 10 $\deg$) was introduced to avoid crowded fields with many stars.

\begin{table*}
	\centering
	\begin{tabular}{||c | c c c c c c c||} 
		\hline
		\textbf{Event} & \textbf{R.A. (J2000)} & \textbf{Dec (J2000)} & \textbf{90\% area} & \textbf{ZTF obs} & \textbf{Latency} & \textbf{Signalness}& \textbf{References}\\
		& \textbf{[deg]}&\textbf{[deg]}& \textbf{[sq. deg.]}& \textbf{[sq. deg.]} & \textbf{[hours]} &\\
		\hline
		IC190503A & 120.28 & +6.35 & 1.9 & 1.4 & 10.2 & 36\% & \cite{ic190503a} \\\ 
		&&&&&&& \cite{ic190503a_ztf} \\ 
		\hline	 IC190619A & 343.26 & +10.73 & 27.2 & 21.6 & 20.9 & 55\% & \cite{ic190619a} \\\ 
		&&&&&&& \cite{ic190619a_ztf} \\ 
		\hline	 IC190730A & 225.79 & +10.47 & 5.4 & 4.5 & 7.5 & 67\% & \cite{ic190730a} \\\ 
		&&&&&&& \cite{ic190730a_ztf} \\ 
		\hline	 IC190922B & 5.76 & $-$1.57 & 4.5 & 4.1 & 8.0 & 51\% & \cite{ic190922b} \\\ 
		&&&&&&& \cite{ic190922b_ztf} \\ 
		\hline	 IC191001A & 314.08 & +12.94 & 25.5 & 23.1 & 7.4 & 59\% & \cite{ic191001a} \\\ 
		&&&&&&& \cite{ic191001a_ztf} \\ 
		\hline	 IC200107A & 148.18 & +35.46 & 7.6 & 6.3 & 2.0 & $-$ & \cite{ic200107a} \\\ 
		&&&&&&& \cite{ic200107a_ztf} \\ 
		\hline	 IC200109A & 164.49 & +11.87 & 22.5 & 22.4 & 32.4 & 77\% & \cite{ic200109a} \\\ 
		&&&&&&& \cite{ic200109a_ztf} \\ 
		\hline	 IC200117A & 116.24 & +29.14 & 2.9 & 2.7 & 22.0 & 38\% & \cite{ic200117a} \\\ 
		&&&&&&& \cite{ic200117a_ztf} \\ 
		&&&&&&& \cite{ic200117a_ztf_2} \\ 
		\hline	 IC200512A & 295.18 & +15.79 & 9.8 & 9.3 & 1.7 & 32\% & \cite{ic200512a} \\\ 
		&&&&&&& \cite{ic200512a_ztf} \\ 
		\hline	 IC200530A & 255.37 & +26.61 & 25.3 & 22.0 & 0.2 & 59\% & \cite{ic200530a} \\\ 
		&&&&&&& \cite{ic200530a_ztf} \\ 
		&&&&&&& \cite{ic200530a_ztf_2} \\ 
		\hline	 IC200620A & 162.11 & +11.95 & 1.7 & 1.2 & 25.8 & 32\% & \cite{ic200620a} \\\ 
		&&&&&&& \cite{ic200620a_ztf} \\ 
		\hline	 IC200916A & 109.78 & +14.36 & 4.2 & 3.6 & 14.7 & 32\% & \cite{ic200916a} \\\ 
		&&&&&&& \cite{ic200916a_ztf} \\ 
		&&&&&&& \cite{ic200916a_ztf_2} \\ 
		\hline	 IC200926A & 96.46 & $-$4.33 & 1.7 & 1.3 & 4.1 & 44\% & \cite{ic200926a} \\\ 
		&&&&&&& \cite{ic200926a_ztf} \\ 
		\hline	 IC200929A & 29.53 & +3.47 & 1.1 & 0.9 & 14.1 & 47\% & \cite{ic200929a} \\\ 
		&&&&&&& \cite{ic200929a_ztf} \\ 
		\hline	 IC201007A & 265.17 & +5.34 & 0.6 & 0.6 & 4.8 & 88\% & \cite{ic201007a} \\\ 
		&&&&&&& \cite{ic201007a_ztf} \\ 
		\hline	 IC201021A & 260.82 & +14.55 & 6.9 & 6.3 & 43.7 & 30\% & \cite{ic201021a} \\\ 
		&&&&&&& \cite{ic201021a_ztf} \\ 
		\hline	 IC201130A & 30.54 & $-$12.10 & 5.4 & 4.5 & 7.1 & 15\% & \cite{ic201130a} \\\ 
		&&&&&&& \cite{ic201130a_ztf} \\ 
		\hline	 IC201209A & 6.86 & $-$9.25 & 4.7 & 3.2 & 16.9 & 19\% & \cite{ic201209a} \\\ 
		&&&&&&& \cite{ic201209a_ztf} \\ 
		\hline	 IC201222A & 206.37 & +13.44 & 1.5 & 1.4 & 35.2 & 53\% & \cite{ic201222a} \\\ 
		&&&&&&& \cite{ic201222a_ztf} \\ 
		\hline	 IC210210A & 206.06 & +4.78 & 2.8 & 2.1 & 0.2 & 65\% & \cite{ic210210a} \\\ 
		&&&&&&& \cite{ic210210a_ztf} \\ 
		\hline	 IC210510A & 268.42 & +3.81 & 4.0 & 3.7 & 5.1 & 28\% & \cite{ic210510a} \\\ 
		&&&&&&& \cite{ic210510a_ztf} \\ 
		\hline	 IC210629A & 340.75 & +12.94 & 6.0 & 4.6 & 15.4 & 35\% & \cite{ic210629a} \\\ 
		&&&&&&& \cite{ic210629a_ztf} \\ 
		\hline	 IC210811A & 270.79 & +25.28 & 3.2 & 2.7 & 26.7 & 66\% & \cite{ic210811a} \\\ 
		&&&&&&& \cite{ic210811a_ztf} \\ 
		\hline	 IC210922A & 60.73 & $-$4.18 & 1.6 & 1.2 & 16.1 & 92\% & \cite{ic210922a} \\\ 
		&&&&&&& \cite{ic210922a_ztf} \\ 
		\hline
	\end{tabular}
	\caption{Summary of the 24 neutrino alerts followed up by ZTF since survey start on 2018 March 20.}
	\label{tab:nu_alerts}
\end{table*}

\begin{figure}
	\centering \includegraphics[width=0.45\textwidth]{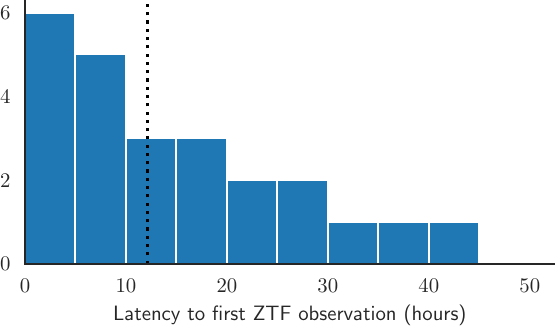}
	\caption{Latency between neutrino detection and first ZTF coverage. The median latency time of 12.2 hours is indicated by the vertical dotted line.}
	\label{fig:ztf_latency}
\end{figure}

Each neutrino localisation region can typically be covered by one or two observations of fields in a predefined ZTF `grid' tiling of the sky. Multiple observations are scheduled for each field, with both $g$ and $r$ filters, and a separation of at least 15 minutes between images. These observations typically last for 300\,s, with a typical limiting magnitude of 21.5.  ToO observations are typically conducted on the first two nights following a neutrino alert, before swapping to serendipitous coverage with shorter 30\,s exposures and a 2-day cadence as part of the public survey \citep{ztf_survey_19}. As can be seen in Figure \ref{fig:ztf_latency}, our first coverage of events has a median latency of 12.2 hours from neutrino detection. Some latency is unavoidable because the neutrino localisation itself is typically only released with a delay of $\gtrsim$2 hours, but additional latency arises primarily due to observability constraints. Poor weather can prevent observations on the first night after neutrino detection, leading to 20\% of alerts observed with a latency >24 hours. Serendipitous coverage from the public survey, with a median latency of 24 hours after neutrino detection, reduces the latency for some campaigns.

As for all ZTF data, these observations are first processed by the Infra-red Processing and Analysis Centre (IPAC) to identify detections in difference images \citep{ztf_data_processing}. These detections are then packaged as `alerts' \citep{zads_19}, and processed by our dedicated data analysis pipeline, \textit{NuZTF} \citep{nuztf}, which searches for extragalactic ZTF detections coincident with external triggers. For neutrinos followed-up by ZTF, we define spatial coincidence as requiring that an object lies within the reported 90\% localisation rectangle from IceCube, and define temporal coincidence as requiring that an object is detected at least once following the neutrino arrival time.

\textit{NuZTF} is built using the \emph{AMPEL} software framework \citep{ampel}, based on a search algorithm for extragalactic transients. Cuts are applied to reject spurious detections, stars and solar system objects (see \citealt{bran} for more details). Searching for detections in the window from neutrino arrival time to 14 days post-neutrino, these cuts typically yield 1 good candidate per $\sim$3 sq. deg. of observed sky.

Promising candidates are prioritised for spectroscopic classification, to confirm or rule out a possible association with a given neutrino. Once classified, an object can then be cross-referenced to relevant neutrino emission scenarios for that population. In particular, optical signatures we look for include:

\begin{itemize}
	\item \textbf{Supernovae with evidence of CSM interaction}. High-energy neutrinos are thought to be produced when CCSNe occur within a dense circumstellar medium (CSM), with the resultant shock collisions then generating neutrino emission simultaneously with the optical lightcurve \citep{murase_csm_sn_11}. The presence of such CSM interaction also results in characteristic narrow lines in the optical spectrum, so these models generally apply to the Type IIn supernova population which exhibits these lines. The neutrino emission is expected to be highest close to optical peak, and to then decay over time. In this case, the expected optical signature would be any supernova with evidence of ongoing CSM interaction.\\
	
	\item \textbf{Supernovae with relativistic jets}. Some supernovae have been observed to launch relativistic jets as part of the core-collapse process \citep{galama_98}. Those jets which proceed to escape the surrounding stellar envelope and CSM can be observed as long GRBs if they are oriented towards Earth. Analogously, where an on-axis supernova jet does not escape the stellar envelope, there would instead be a so-called `choked jet' \citep{nakar_15_llgrb}. For both scenarios, neutrino emission would primarily be expected during the `prompt phase', in the $\sim$100s after supernova explosion \citep{waxman_bahcall_97_grb, senno_choked_jets_16}. This scenario would then lead to a young supernova, typically of Type Ic-BL, appearing at the location of the neutrino. The supernova would have an explosion time compatible with the neutrino detection time, and since SNe brighten over a period of days, this optical signature would be delayed relative to the neutrino itself.\\
	
	\item \textbf{GRB Afterglows.} Another signature of the supernova jet scenario would be the direct detection of a long-GRB afterglow. Models have also predicted neutrino emission for short GRBs, so a short-GRB afterglow could also be a potential counterpart \citep{waxman_bahcall_97_grb}. These GRB afterglows would not be detected before the neutrino detection, and would fade rapidly over the next few hours before falling below the ZTF detection threshold.\\
	
	\item \textbf{AGN Flares}. AGN flares, and especially blazar flares, have been suggested as neutrino sources \citep{bednarek_99}, though the neutrino emission itself would not necessarily be directly correlated to the optical emission. For example, for the standard two-hump Spectral Energy Distribution (SED) model, the optical emission could serve primarily as a tracer for photon target density but not necessarily PeV proton luminosity. We restrict ourselves to searches for AGN undergoing significant optical flaring coincident with a neutrino. Neutrinos could also be produced in AGN without coincident optical flares, but such neutrino emission scenarios are not best probed with an optical follow-up program such as ours.\\
	
	\item \textbf{Tidal Disruption Events}. TDEs have been suggested as neutrino sources, through multiple emission channels such as jets, outflows or in coronae (see \citealt{hayasaki_21} for a recent review). The timescale for neutrino production remains unclear, but would not be expected prior to the TDE itself. Non-thermal emission from TDEs can last several hundred days, so the signature in this case would be any `ongoing' TDE coincident with a neutrino.\\
\end{itemize}

We do not enforce any additional cut on candidate distance, because IceCube detects neutrinos emitted throughout the universe rather than being restricted only to the local universe (see \citealt{strotjohann_19} for a more detailed explanation of this effect). However, given the limiting magnitude of ZTF, the candidates we find will nonetheless be biased towards lower redshifts.

We do not explicitly reject objects with a history of variability, because variable objects have been proposed as possible neutrino sources. However, our program is intended to identify increased optical flux that is contemporaneous with a neutrino's detection, so only variable objects with significantly enhanced flux relative to reference images are selected by our pipeline. The blazar flare of TXS 0506+056 fell into this category \citep{ic_txs_mm_18}, and we would be capable of identifying similar examples. 

To date, the \textit{NuZTF} pipeline has identified 172 candidates for visual inspection out of an observed area of 154.33 sq. deg across 24 neutrinos, using a search window of 14 days after each neutrino detection. This corresponds to an initial density of 1.05 candidates per sq. deg. of sky. The full list of candidates for each neutrino is given in the Appendix. 

Visual inspection then enables us to further classify objects and reject background detections. Viewing difference images directly enables us to identify additional image artefacts. We select likely stars through cross-matches to Gaia \citep{gaia_dr2}, where we reject sources with significant (3$\sigma$) evidence for parallax, and to SDSS star/galaxy morphology classifications \citep{sdsss_edr_02}. 

\begin{figure}
	\centering \includegraphics[width=0.45\textwidth]{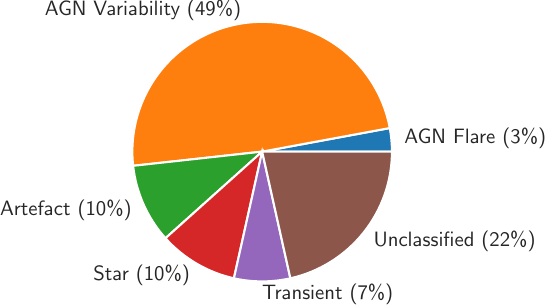}
	\caption{Breakdown of the classification of 172  candidates selected by our program for visual inspection.}
	\label{fig:candidates}
\end{figure}

We then flag AGN through matches to catalogued sources in the Milliquas catalogue \citep{milliquas}, or via WISE colour cuts \citep{wise_10, stern_12}. We further cross-match to NASA's NED database, to flag any missing catalogued sources \citep{ned_91}. We seek to distinguish between `routine' AGN variability and extreme AGN flares. We search for evidence of flaring activity at the time of neutrino detection using the data provided in the ZTF alert packets \citep{zads_19}, which are based on difference images. For cases where a source appears to be significantly variable, or may have been flaring at the time of neutrino detection, we run dedicated forced photometry on the science images to produce a source lightcurve \citep{ztf_data_processing}. We reject AGN with no evidence for contemporaneous flaring as `AGN variability'. After removing those sources flagged as stars (17), image artefacts (17) or AGN variability (84), we are left with 54 `interesting candidates' (5 AGN flares, 12 confirmed transients and 37 unclassified sources). The full breakdown in classification is shown in Figure \ref{fig:candidates}. 

\begin{figure}
	\centering \includegraphics[width=0.45\textwidth]{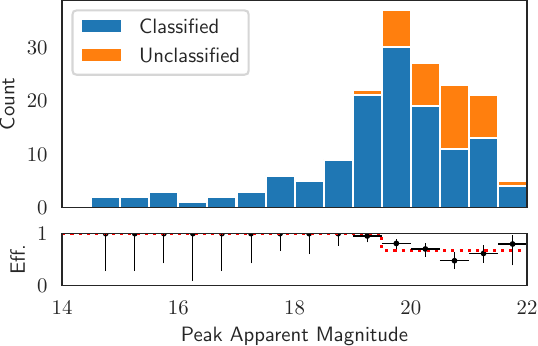}
	\caption{Top: Apparent magnitude distribution of candidates selected for visual inspection. Bottom: Classification efficiency as a function of peak apparent magnitude. The red dashed line indicates our step-function approximation of classification efficiency.}
	\label{fig:completeness}
\end{figure}

These interesting candidates include potential transients, which we seek to classify spectroscopically. Some objects will have already been classified serendipitously, in particular those brighter than 19.0 mag selected by the ZTF Bright Transient Survey \citep{ztf_bts_1, ztf_bts_2}. The efficiency with which candidates were classified can be seen in Figure \ref{fig:completeness}. Above a peak apparent magnitude of 19.5, almost all candidates are classified. There were 106 fainter candidates in total, of which 68\% were classified. The spectroscopic programs which supported our program are listed in Table \ref{tab:spectroscopy}.  

\begin{table}
	\centering
	\begin{tabular}{||c | c  |} 
		\hline
		\textbf{Instrument}  & \textbf{Programs}\\
		\hline
		 SEDm& 2018,  2019, 2020, 2021\\
		 \hline
		 NOT	& 2021B (OPT21B$\_$50, PI: Franckowiak)\\
		 & 2021B (P64-112)\\ 
		 & 2021B (P61-501)\\
		 & 2022A (22A013, PI: Franckowiak)\\
		 \hline
		TNG	&2021B (OPT21B$\_$50, PI: Franckowiak)\\
		& 2022A (22A01, PI: Franckowiak)\\
		\hline
		GEMINI	&2021A  (GN-2021A-Q-116, PI: Kasliwal)\\
		& 2021B (GN-2021B-Q-117, PI: Kasliwal)\\
%		& 2022A (expected in Dec)\\ 
		\hline
		GTC &2020B (GTC73-20B, PI: Amaro Seoane) \\
		\hline
	\end{tabular}
	\caption{Summary of dedicated spectroscopic programs for our neutrino follow-up program.}
	\label{tab:spectroscopy}
\end{table}

\begin{figure}
	\centering \includegraphics[width=0.45\textwidth]{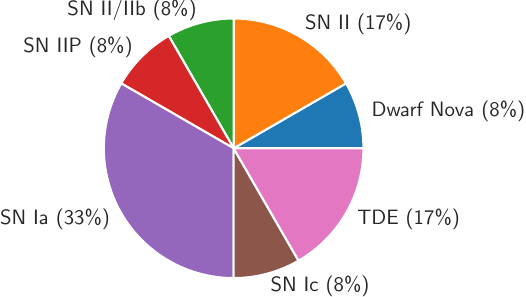}
	\caption{Breakdown of the 12 identified transients by subclass.}
	\label{fig:transient_pie}
\end{figure}

The transients are further broken down by subclass in Figure \ref{fig:transient_pie}. Four could be immediately excluded as candidates based on their classification as SNe Ia, a population not predicted to emit high-energy neutrinos. Of the remainder,  beyond the two TDEs, no further sources exhibited electromagnetic signatures from the theoretically-predicted neutrino emission scenarios listed above. However, we cannot exclude the possibility that future theoretical work proposes additional neutrino emission scenarios not considered here, and therefore we cannot definitively rule out these transients as neutrino sources. We can state that none of them are consistent with existing models.

A selection of highlighted results is given in the following sections. We specifically list transients which were unclassified at the time of our ToO observations, and the additional follow-up that we took to confirm their nature. ZTF data for two other candidate neutrino sources from the literature, PKS 1502+106 and BZB J0955+3551, are also outlined in Section \ref{sec:literature}. We omit ZTF data for the probable neutrino-TDEs AT 2019dsg and AT 2019fdr, as these have already been released in dedicated publications \citep{bran, tywin}. All other classified sources are listed in Appendix tables \ref{tab:ic190503a}-\ref{tab:ic210811a}. 

For four neutrino campaigns (IC200107A, IC201007A, IC201222A and IC210922A), no candidates were identified, and there are no corresponding lists in the appendix.

\section{Candidate Transient Counterparts}
\label{sec:transients}

\subsection{SN 2019pqh and IC190922B}
\label{sec:SN 2019pqh}
\begin{figure}
	\centering \includegraphics[width=0.48\textwidth]{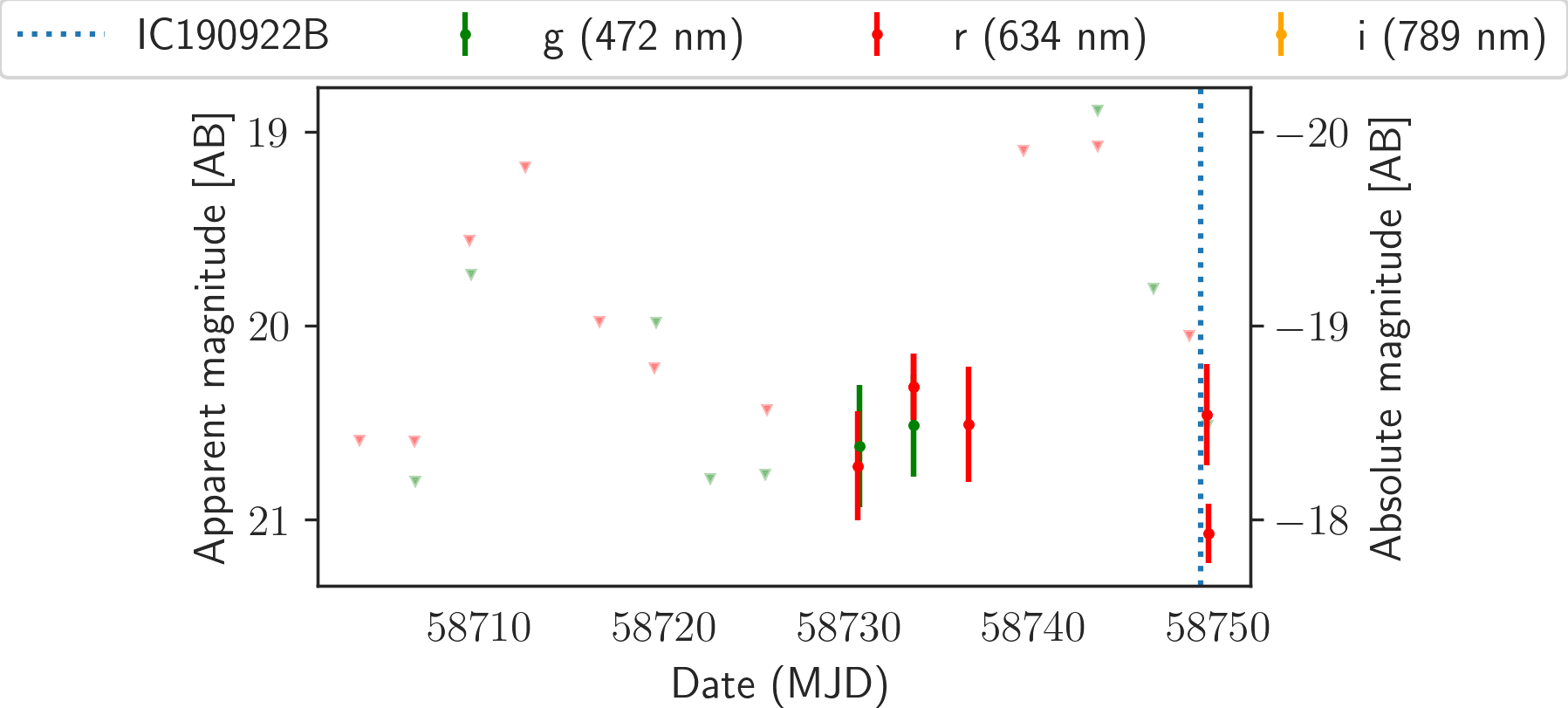}
	\caption{ZTF lightcurve of SN 2019pqh. The arrival time of neutrino IC190922B is marked by the dashed blue line.}
	\label{fig:SN2019pqh_ligtcurve}
\end{figure}

Follow-up of IC190922B by ZTF identified the candidate supernova SN 2019pqh/ZTF19abxtupj \citep{ic190922b_ztf}. The lightcurve is shown in Figure \ref{fig:SN2019pqh_ligtcurve}, where upper limits are illustrated with triangles. The arrival time of the neutrino on 2019 September 22 is marked with a dotted line, and the supernova is detected in the subsequent ToO observations. The neutrino arrival time was close to optical peak, consistent with a CSM-interaction scenario.

\begin{figure}
	\centering \includegraphics[width=0.45\textwidth]{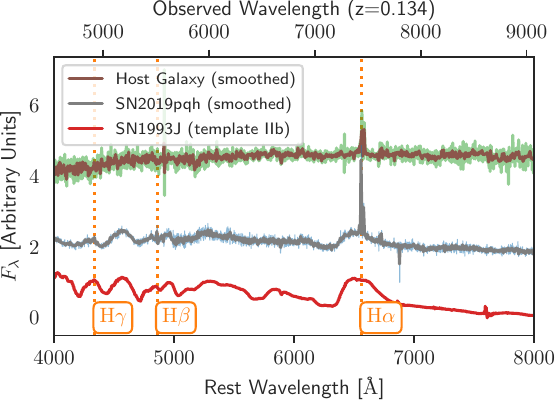}
	\caption{Spectrum of SN 2019pqh, taken on 2019 September 28. A historical spectrum of the host galaxy taken by SDSS, and a similar spectrum of a Type IIb supernova, are provided for comparison.}
	\label{fig:SN2019pqh_spectrum}
\end{figure}

However, a spectrum was taken by the \emph{NUTS2 collaboration} \citep{nuts2_18}, and the supernova was classified as a \hbox{Type II} supernova without spectroscopic signatures of CSM interaction \citep{sn2019pqh_class}. A higher-resolution spectrum of the object was also obtained on 2019 September 28, shown in Figure \ref{fig:SN2019pqh_spectrum}, using the \emph{Low Resolution Imaging Spectrometer} (LRIS) spectrograph at the Keck observatory (PI: Yan) \citep{keck_lris_95}. A historical spectrum of the host galaxy, taken by the \emph{Sloan Digital Sky Survey}  (SDSS; \cite{sdss_dr14}), is also shown in Figure \ref{fig:SN2019pqh_spectrum}. Both the transient and host galaxy exhibit prominent Balmer lines, highlighted in orange in Figure \ref{fig:SN2019pqh_spectrum}, from which a redshift of 0.134 is derived. A template-matching classification using SNID \citep{snid_07} confirms a \hbox{Type II} supernova classification, with the best match being a Type IIb supernova (SN 1993J, \citealt{sn1993j_95}) 2 days before peak, also shown in Figure \ref{fig:SN2019pqh_spectrum}. 

With this redshift, a peak absolute magnitude of $-18.6$ was derived, atypically bright for such a Type II supernova (see e.g. \citealt{lyman_16}). One explanation for this enhanced luminosity could be CSM interaction, through which additional kinetic energy is converted to electromagnetic emission. However, the lack of corresponding narrow line spectroscopic signatures generally disfavours the existence of CSM-interaction, and thus any associated neutrino emission from this object. It is therefore likely that SN 2019pqh is instead unrelated to the neutrino IC190922B.

\subsection{SN 2020lam and IC200530A}

\begin{figure}
	\centering
	\includegraphics[width=0.45\textwidth]{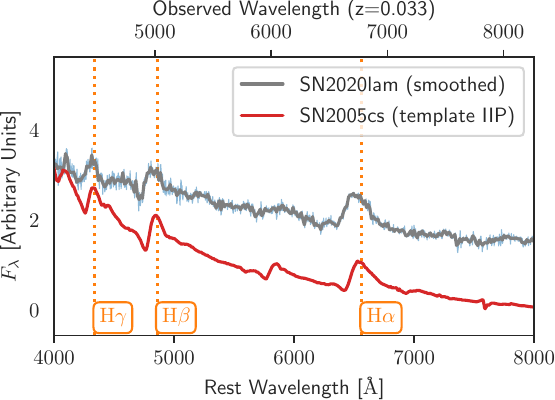}
	\caption{Spectrum of SN 2020lam, taken on 2020 June 6. A similar spectrum, from Type IIP supernova SN 2005cs, is shown for comparison.}
	\label{fig:SN2020lam_spectrum}
\end{figure}

ZTF serendipitously observed the localisation of neutrino alert IC200530A on 2020 May 30, just 10 minutes after detection \citep{ic200530a}, as part of routine survey operations \citep{ic200530a_ztf}. Additional ToO observations were then conducted on 2020 May 31 in $g$ and $r$ band, and again on 2020 June 1. During ZTF follow-up of IC200530A, SN 2020lam/ZTF20abbpkpa was identified as a candidate supernova and potential optical counterpart \citep{ic200530a_ztf}. Spectroscopic observations were triggered using the NOT/ALFOSC spectrograph on 2020 June 6 (PI: Sollerman), which confirmed SN 2020lam as a Type II supernova \citep{ic200530a_ztf_2} using SNID. This spectrum is shown in Figure \ref{fig:SN2020lam_spectrum}, alongside the matching Type IIP supernova (SN 2005cs, \citealt{sn2005cs_06}) mapped to the same redshift.

\begin{figure}
	\centering \includegraphics[width=0.48\textwidth]{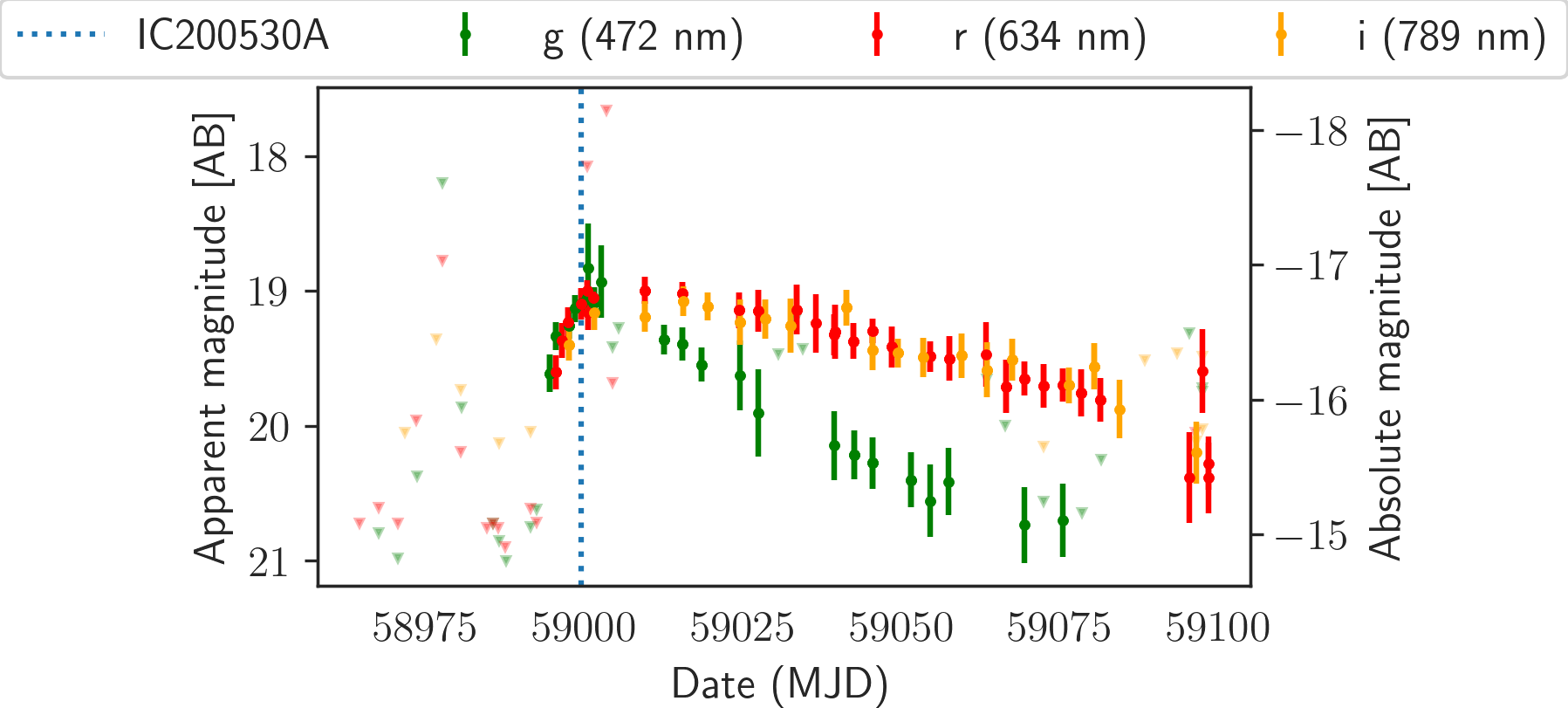}
	\caption{ZTF lightcurve of SN 2020lam. The arrival time of neutrino IC200530A is marked by the dashed blue line.}
	\label{fig:SN2020lam_lightcurve}
\end{figure}

As seen in the lightcurve in Figure \ref{fig:SN2020lam_lightcurve}, the supernova was close to peak at neutrino detection time.  The object then rapidly cooled, and thus reddened, as is typical for supernovae. Given the neutrino arrival time, CSM-interaction would be the only viable neutrino production mechanism. However, the spectrum shown in Figure \ref{fig:SN2020lam_spectrum} had no narrow lines, and therefore did not provide any evidence supporting such CSM interaction. SN 2020lam was therefore likely unrelated to IC200530A.

\subsection{SN 2020lls and IC200530A}

\begin{figure}
	\centering \includegraphics[width=0.45\textwidth]{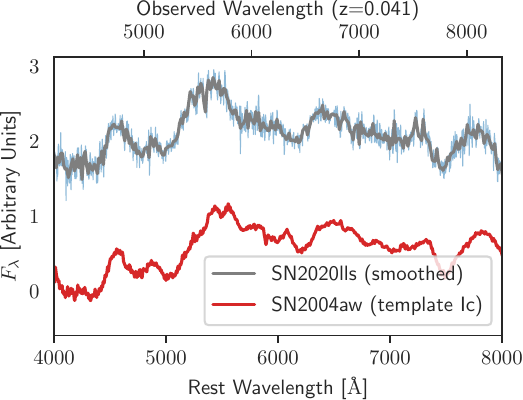}
	\caption{Spectrum of SN 2019lls, taken on 2020 June 13. A similar spectrum, of Type Ic supernova SN 2004aw, is shown for comparison.}
	\label{fig:SN2020lls_spectrum}
\end{figure}

SN 2020lls/ZTF20abdnpdo was also identified as a candidate supernova on 2020 May 30, during ZTF follow-up of IC200530A  \citep{ic200530a_ztf}. Spectroscopic observations were again triggered using the NOT/ALFOSC spectrograph on 2020 June 12 (PI: Sollerman), which confirmed that SN 2020lls was a Type Ic supernova without broad-line features \citep{ic200530a_ztf_3}. This spectrum is illustrated in Figure \ref{fig:SN2020lls_spectrum}, alongside a matching Type Ic supernova spectrum from SNID mapped to the same redshift \citep{sn2004aw_06}. Given that the supernova had not been detected in alert data prior to the neutrino arrival time, and that it belonged to the subpopulation associated with relativistic jets, SN 2020lls was a candidate for the choked-jet neutrino production model.

\begin{figure}
	\centering \includegraphics[width=0.49\textwidth]{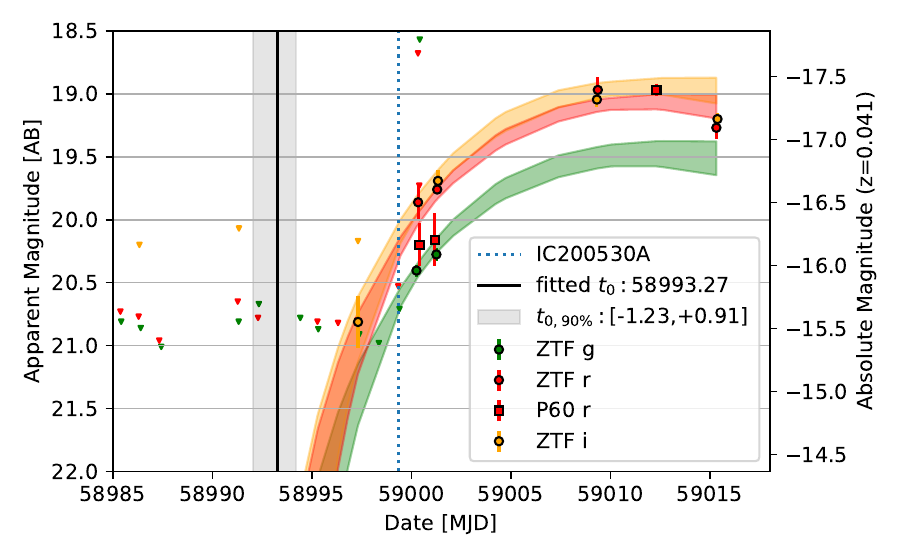}
	\caption{ZTF lightcurve of SN 2020lls. The arrival time of neutrino IC200530A is marked with the blue dotted line. The supernova model fit from \emph{MOSFIT} is indicated by the shaded orange/red/green bands, and the the best-fit explosion time is given by the vertical black line.}
	\label{fig:ztf20abdnpdo}
\end{figure}

However, as can be seen in Figure \ref{fig:ztf20abdnpdo}, forced photometry analysis \citep{ztffps} revealed a lower-threshold $i$-band ZTF detection preceding the neutrino arrival. Additionally, modelling of the lightcurve using the \emph{MOSFIT} software \citep{mosfit_18} revealed an estimated explosion date predating the neutrino by a week. In combination, these results disfavoured any supernova explosion origin for the neutrino, suggesting that SN 2020lls was instead unrelated to IC200530A \citep{ic200530a_ztf_3}. 

\section{AGN flare candidates}
\label{sec:AGN flares}

\begin{figure} 
	\centering \includegraphics[width=0.48\textwidth]{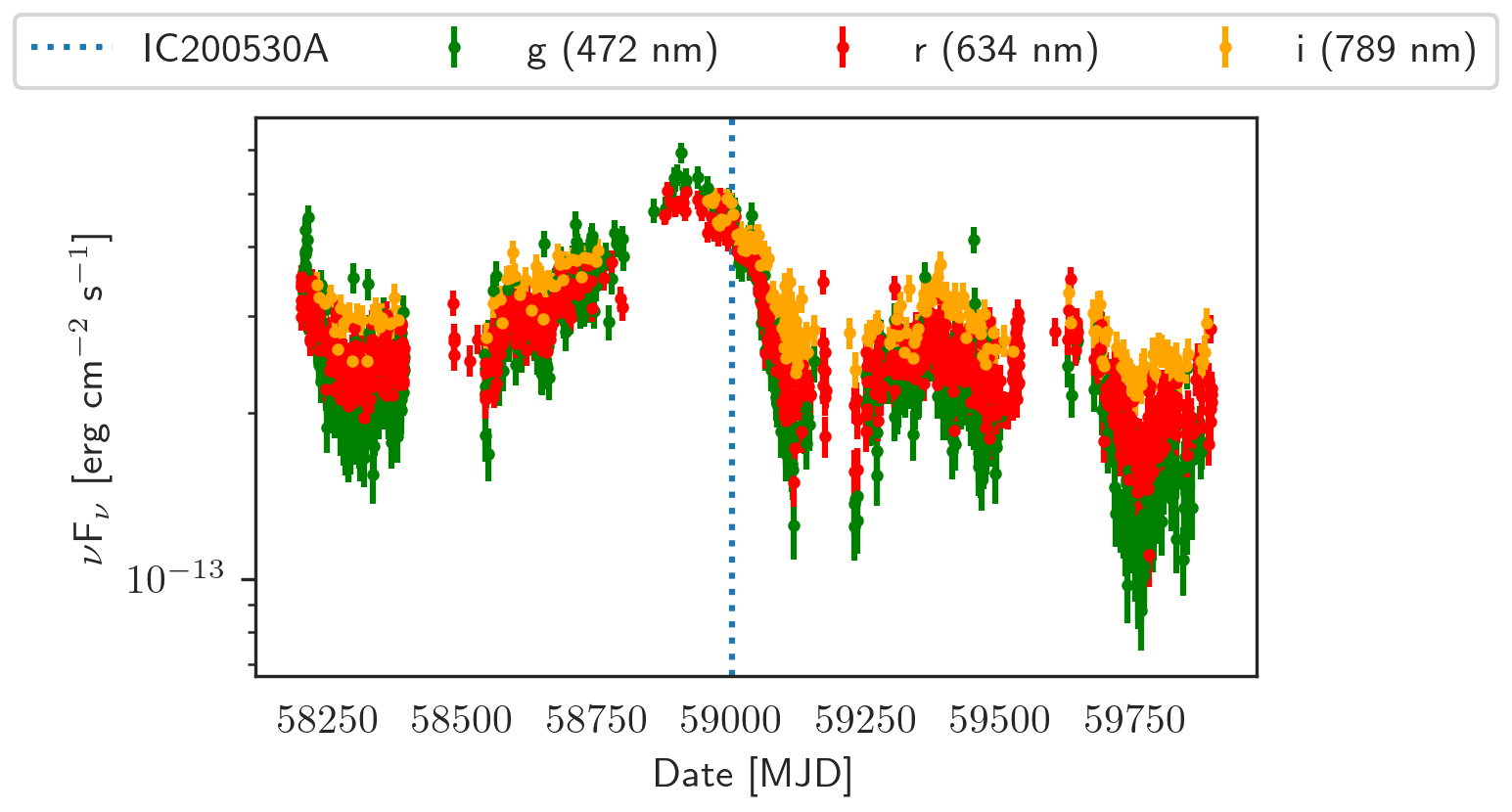}
%	\caption{ZTF lightcurve of likely QSO ZTF18aavecmo, with coincident high-energy neutrino IC200530A illustrated with the vertical dotted line.}
		\centering \includegraphics[width=0.48\textwidth]{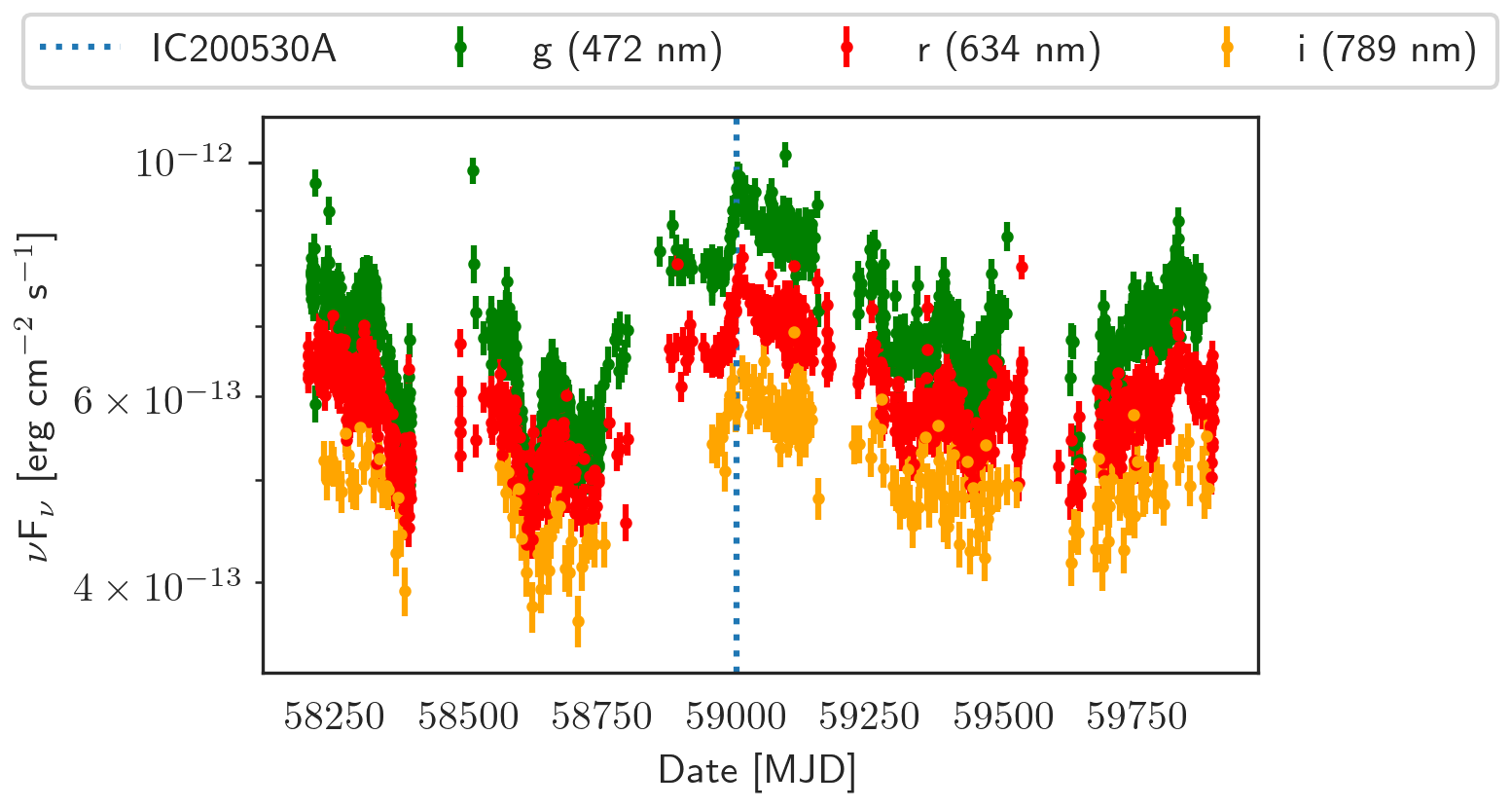}
		\centering \includegraphics[width=0.48\textwidth]{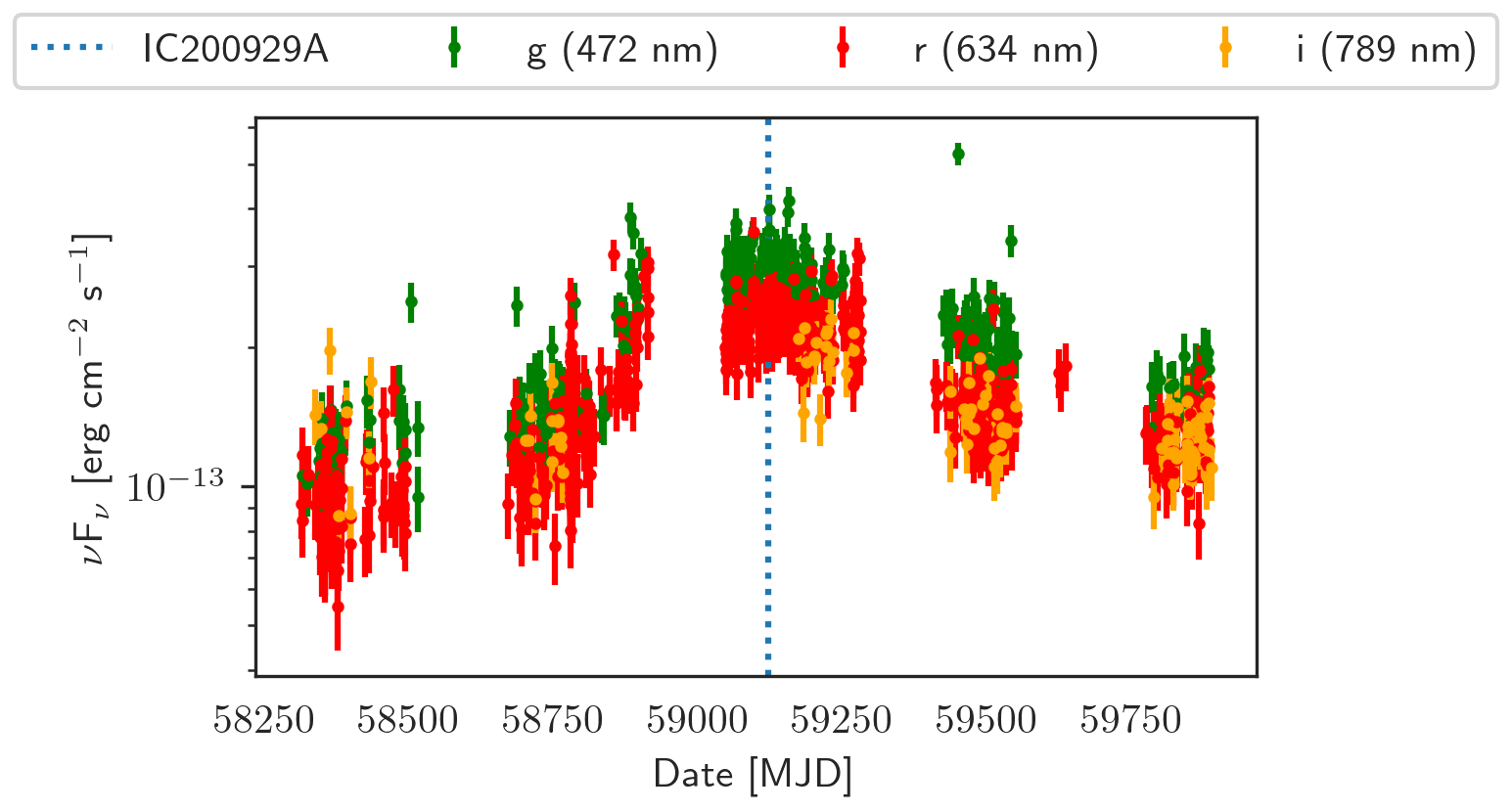}
		\centering \includegraphics[width=0.48\textwidth]{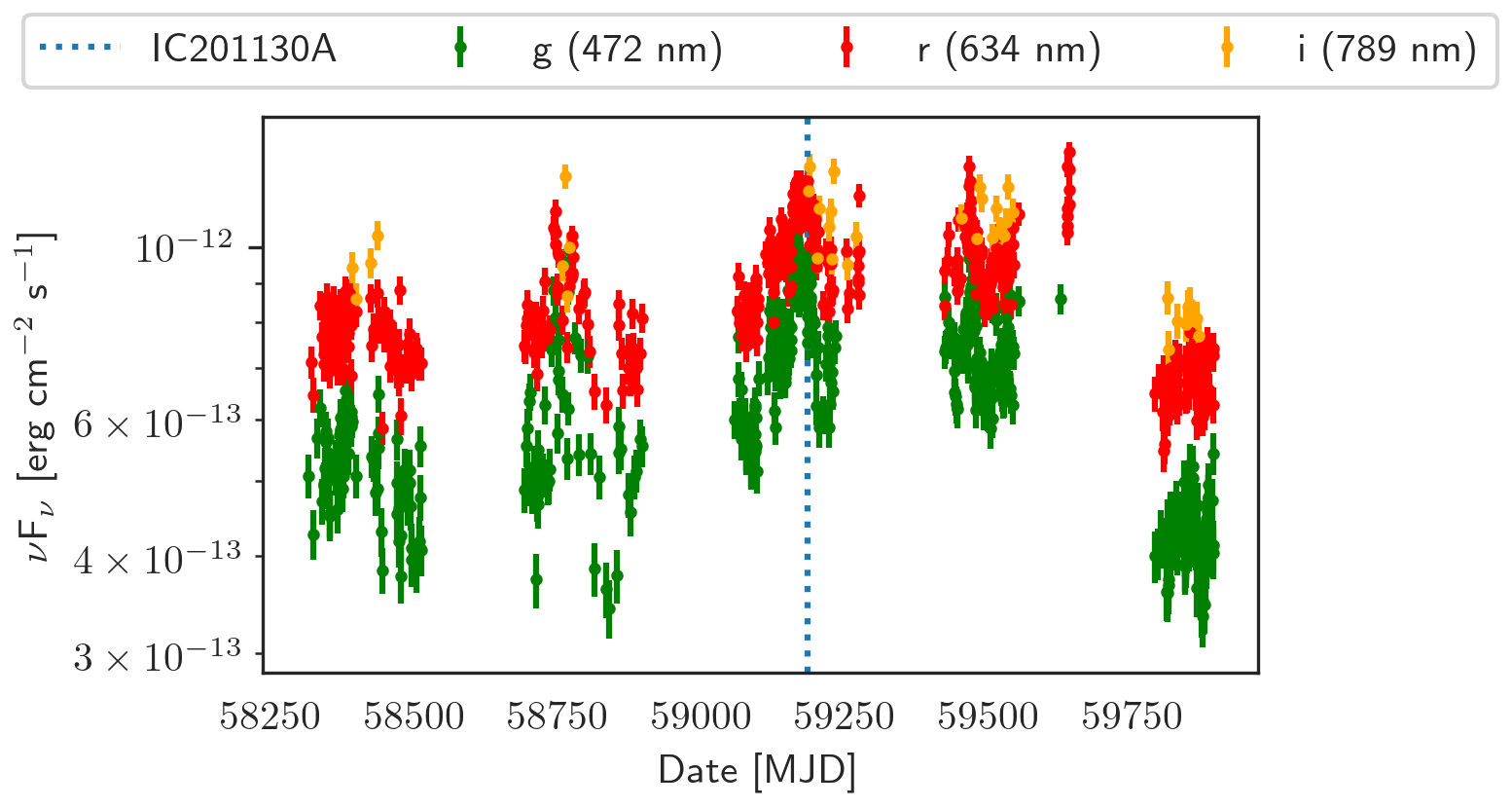}
		\centering \includegraphics[width=0.48\textwidth]{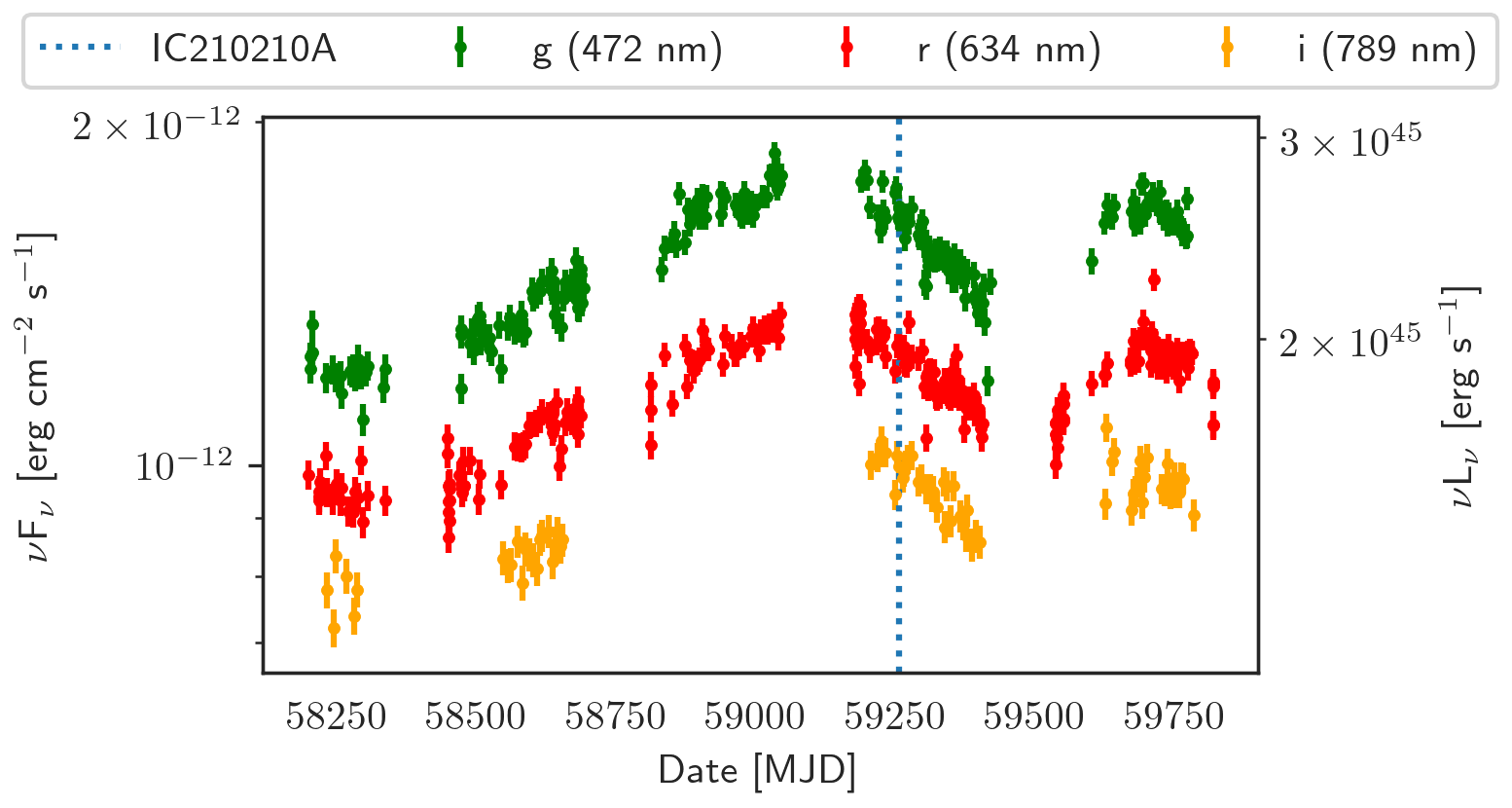}
		\caption{ZTF lightcurve of 5 AGN flares coincident with high-energy neutrinos. From top to bottom, the sources are: ZTF18aavecmo, ZTF18abrwqpr, ZTF20aamoxyt, ZTF18abxrpgu, ZTF19aasfvqm.}
	\label{fig:flares}
\end{figure}

While the vast majority of AGN detections from our pipeline were categorised as `AGN variability', visual inspection revealed five AGN which appeared to possibly undergo optical flaring at the time of neutrino detection. The forced photometry lightcurves of these five flares are shown in Figure \ref{fig:flares}. We attempt to quantify whether the optical lightcurves of these AGN identify them as candidate neutrino sources. 

We can consider possible optical signatures associated with neutrino emission. One scenario is the optical flaring observed for TXS 0506+056 during the detection of neutrino IC170922A \citep{ic_txs_mm_18}. In particular, the optical apparent V-band magnitude of TXS 0506+056 was observed to increase from 15.0 to 14.5 during the time of neutrino detection, corresponding to a flux increase of  >50\%, over a period of $50$ days, relative to the pre-neutrino baseline. 

AGN can also exhibit short-term variability for periods of hours or days, but we caution that the detection of a high-energy neutrino alert is a process that requires a substantial fluence at the IceCube detector, even after accounting for the significant Eddington bias associated with cosmic neutrino detection \citep{strotjohann_19}. The corresponding neutrino flux that is required is inversely proportional to the duration of neutrino emission, and therefore associating a neutrino detection with a temporary electromagnetic signature lasting hours or days would imply an extremely high average neutrino flux for the duration of that signature. Such highly luminous rapid neutrino flares are not well motivated theoretically, it is therefore unlikely that short AGN flares are indicators of neutrino production. 

In contrast, longer-term electromagnetic signatures can serve as tracers for neutrino emission. For example, month-long flaring periods of substantially elevated flux can dominate the neutrino emission of blazars (see e.g. \citealt{rodrigues_21}). Very long flares, with durations of years, could also be relevant for neutrino production. However, given the relatively short baseline of ZTF observations, our neutrino follow-up program is not well-suited to identify them. We therefore restrict ourselves to searching for such month-long optical flares, as was observed for TXS 0506+056. 

We calculate the median flux for each of the five AGN, and each ZTF filter, in a $\pm$25 day window centered on the neutrino detection. We divide this instantaneous flux by the median flux of the source in that filter over the entire $\sim$4 year ZTF baseline, giving a proxy for relative optical flare strength. These values are given in Table \ref{tab:agn_flares}. Of the five AGN, only two (ZTF18aavecmo and ZTF20aamoxyt) had a median instantaneous flux >50\% above the baseline median flux. ZTF18aavecmo reached this threshold in both g and r band, while ZTF20aamoxyt reached this threshold only in r. We conclude that the remaining three AGN (ZTF18abrwqpr, ZTF18abxrpgu, ZTF19aasfvqm) do not exhibit substantial neutrino-coincident optical flares, and we therefore find no evidence to suggest they are counterparts to high-energy neutrinos.

\begin{table*}
	\centering
	\begin{tabular}{||c | c | c c | c |} 
		\hline
		\textbf{Object} & \textbf{Filter} &\textbf{Inst. Flux} & \textbf{Med. flux} & \textbf{Inst. Flux} / \textbf{Med. flux} \\
		&& [10$^{-13}$ erg cm$^{-2}$ s$^{-1}$] & [10$^{-13}$ erg cm$^{-2}$ s$^{-1}$] & \\
		\hline
		ZTF18aavecmo & g & 4.5 & 2.5 & 1.82 \\ 
		ZTF18aavecmo & r & 4.3 & 2.6 & 1.67 \\ 
		ZTF18aavecmo & i & 4.5 & 3.2 & 1.40 \\ 
		\hline 
		ZTF18abrwqpr & g & 9.0 & 6.9 & 1.31 \\ 
		ZTF18abrwqpr & r & 7.2 & 5.8 & 1.24 \\ 
		ZTF18abrwqpr & i & 6.0 & 5.0 & 1.21 \\ 
		\hline 
		ZTF20aamoxyt & g & 3.2 & 2.3 & 1.38 \\ 
		ZTF20aamoxyt & r & 2.4 & 1.6 & 1.51 \\ 
		\hline 
		ZTF18abxrpgu & g & 8.9 & 6.7 & 1.33 \\ 
		ZTF18abxrpgu & r & 11.2 & 8.9 & 1.27 \\ 
		\hline 
		ZTF19aasfvqm & g & 16.5 & 14.7 & 1.12 \\ 
		ZTF19aasfvqm & r & 12.6 & 11.6 & 1.08 \\ 
		ZTF19aasfvqm & i & 10.0 & 8.8 & 1.14 \\ 
		\hline 
		
	\end{tabular}
	\caption{Summary of the 5 AGN flares coincident with neutrinos, including the instantaneous flux during neutrino detection, median flux over the entire ZTF baseline, and the ratio of these values.}
	\label{tab:agn_flares}
\end{table*}

ZTF18aavecmo (top panel of Figure \ref{fig:flares}), cross-matched to source WISEA J170539.32+273641.2, is classified as a likely QSO in the Milliquas catalogue. It underwent a single coherent flare lasting approximately one year, with a peak flux roughly triple the quiescent flux measured by ZTF. It was coincident with neutrino IC200530A, detected during the decay of the optical flare. However, this flare was extremely faint, with a median flux at the time of neutrino detection was $\nu F_{\nu} \approx 5 \times 10^{-13}$ erg cm$^{-2}$ s$^{-1}$.  This is a factor of 20 lower than the flux observed for TXS 0506+056 during the detection of IC170922A \citep{ic_txs_mm_18}. 

ZTF20aamoxyt (middle panel of Figure \ref{fig:flares}) is a likely AGN flare, detected coincident with IC200929A. It appears to be spatially consistent with its host galaxy nucleus, and cross-matched to WISEA J015853.53+035126.6. Based on a WISE colour of W1$-$W2=0.7, it is a possible AGN. The neutrino IC200929A was detected during an extended year-long flare. Much like ZTF18aavecmo,  ZTF20aamoxyt at the time of neutrino detection was substantially fainter than TXS 0506+056, with a median flux of $\nu F_{\nu} \approx 3 \times 10^{-13}$ erg cm$^{-2}$ s$^{-1}$. 

We thus identify no optical AGN flares which resemble the multi-wavelength flare of TXS 0506+056 in 2017, from any of our 24 neutrino follow-up campaigns. We emphasize that our results do not preclude a significant degree of neutrino emission from AGN more broadly, but they do disfavour scenarios where the vast majority of astrophysical neutrinos are produced by bright AGN optical flares. There is no tension with scenarios where AGN neutrino emission is not dominated by bright optical flares, for example the `steady state' AGN neutrino models tested in \cite{ic_agn_21} or scenarios where AGN neutrino emission is correlated only to gamma-ray flares. Constraining such scenarios would require more comprehensive multi-wavelength analysis of AGN  lightcurves, incorporating other wavelengths in addition to the optical data presented here. A more systematic study of correlations between ZTF-detected AGN flares and neutrinos, including corresponding chance coincidence probabilities, will be the subject of a future analysis. 

%We further caution that the probability of chance alignment of AGN flares with neutrinos is unlikely to be negligible. 

%ZTF18abrwqpr (upper middle panel of Figure \ref{fig:flares}), cross-matched to source WISEA J165707.06+234643.8, is another likely QSO detected coincident with IC200530A. The lightcurve appears to have undergone an extended year-long flare prior to the detection of the neutrino, with additional short-term variation coincident with the neutrino detection.
%
%ZTF20aamoxyt (middle panel of Figure \ref{fig:flares}) is a likely AGN flare, detected coincident with IC200929A. It appears to be spatially consistent with its host galaxy nucleus, and cross-matched to WISEA J015853.53+035126.6. Based on a WISE colour of W1$-$W2=0.7, it is a possible AGN. The neutrino IC200929A was detected during an extended year-long flare. 
%
%ZTF18abxrpgu (lower middle panel of Figure \ref{fig:flares}) is a likely AGN flare, cross-matched to WISEA J020128.20-121946.2. The WISE colour of W1$-$W2=0.6 supports a possible AGN classification. The source had a history of variability, but was in a somewhat elevated flaring state at the time of neutrino detection.
%
%ZTF19aasfvqm (lower panel of Figure \ref{fig:flares}) was a slowly-evolving flare that lasted approximately three years, with the peak occuring shortly before the neutrino detection. It is crossmatched to a Milliquas quasar, SDSS J134034.75+045241.3.

\section{Candidate neutrino sources from the literature}
\label{sec:literature}

We here provide data on two candidate neutrino sources reported in the literature. However, we caution that none of the objects presented here were selected by our pipeline as ZTF candidates, and therefore are not considered part of our systematic search for neutrino counterparts. We would not claim any such object as a candidate neutrino sources in our neutrino follow-up program, because the chance coincidence probability would be unquantifiable. Any search for additional candidate neutrino sources, beyond those candidates found by our pipeline following ToO observations, would require an independent and unbiased systematic analysis procedure.

\subsection{PKS 1502+106}
The neutrino IC190730A was reported by IceCube in spatial coincidence with PKS 1502+106 \citep{ic190730a}, a particularly gamma-bright Flat Spectrum Radio Quasar (FSRQ)  with a catalogued redshift of z=1.84 \citep{sdss_dr13} . The object was observed by ZTF as part of ToO observations, and was detected under the ZTF candidate name ZTF18aaqnqzx \citep{ic190730a_ztf}.  The blazar had already been repeatedly detected as part of the routine survey operations, with both positive and negative flux changes relative to survey reference images. 

\begin{figure}
	\centering \includegraphics[width=0.48\textwidth]{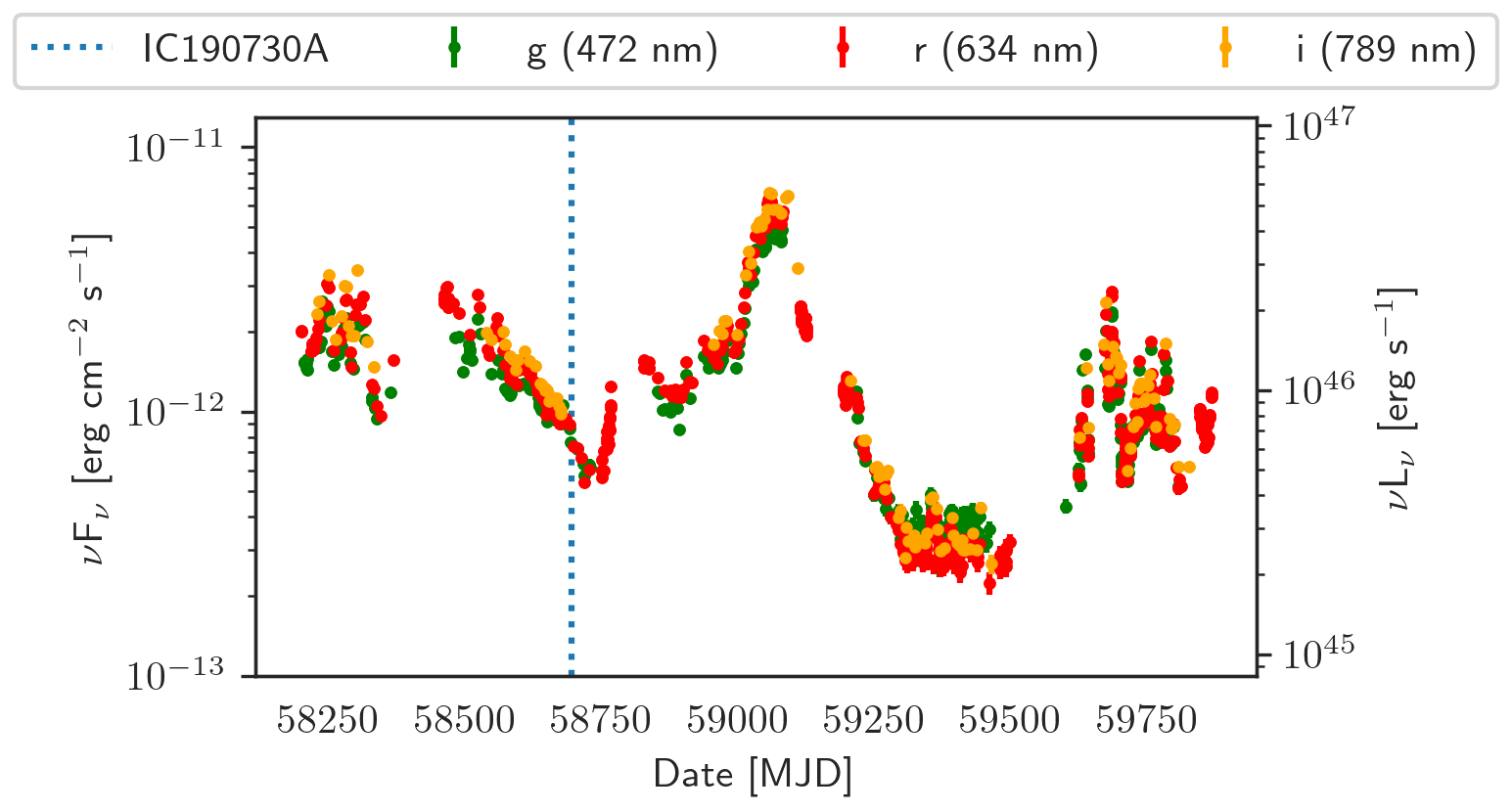}
	\caption{ZTF lightcurve of blazar PKS 1502+106. The arrival time of neutrino IC190730A is marked with the vertical dashed line.}
	\label{fig:pks1502_ligtcurve}
\end{figure}

The blazar lightcurve is shown in Figure \ref{fig:pks1502_ligtcurve}, using data from science images with the ZTF forced photometry service \citep{ztf_data_processing}. The neutrino arrival time is marked in blue. There was no significant flaring observed for this source coincident with the neutrino. The blazar at this point was dimmer than survey reference images, with the neutrino arriving during a year-long fading, and consequently was not selected by our follow-up pipeline as a possible counterpart. There is thus no evidence from the contemporaneous ZTF data to suggest a causal connection between IC190730A and PKS 1502+106, consistent with data from other observatories which did not see any evidence of short-term flaring \citep{franckowiak_20}. 

Data from the Owens Valley Radio Observatory (OVRO) did reveal that the radio flux was elevated in the months preceding the neutrino detection relative to the decade-long observation baseline, behaviour which has also been claimed for TXS 0506+056 and other neutrino-coincident blazars \citep{ovro_19}. Comprehensive time-dependent modelling has found that the detection of a neutrino alert from PKS 1502+106 is consistent with the multi-wavelength observations of this object, so a neutrino-blazar association is plausible but likely unrelated to the flaring activity \citep{rodrigues_21, oikonomou_21}. In any case, the new optical data presented here can be used to further constrain such neutrino emission scenarios.

\subsection{BZB J0955+3551}
IC200107A was a high-energy neutrino reported by IceCube \citep{ic200107a} which was later identified to be in spatial and temporal coincidence with a blazar undergoing a dramatic simultaneous X-ray flare \citep{krauss_ic200107a, giommi_ic200107a}. The source BZB J0955+3551 (also known as 4FGL J0955.1+3551 and 3HSP J095507.9+355101), located at a redshift of 0.55703 \citep{paliya_20}, belongs to the specific subclass of extreme blazars, which are characterised by synchrotron peaks at very high frequencies, which had been proposed as especially promising candidates of high-energy neutrinos \citep{padovani_16}. 

\begin{figure}
	\centering \includegraphics[width=0.48\textwidth]{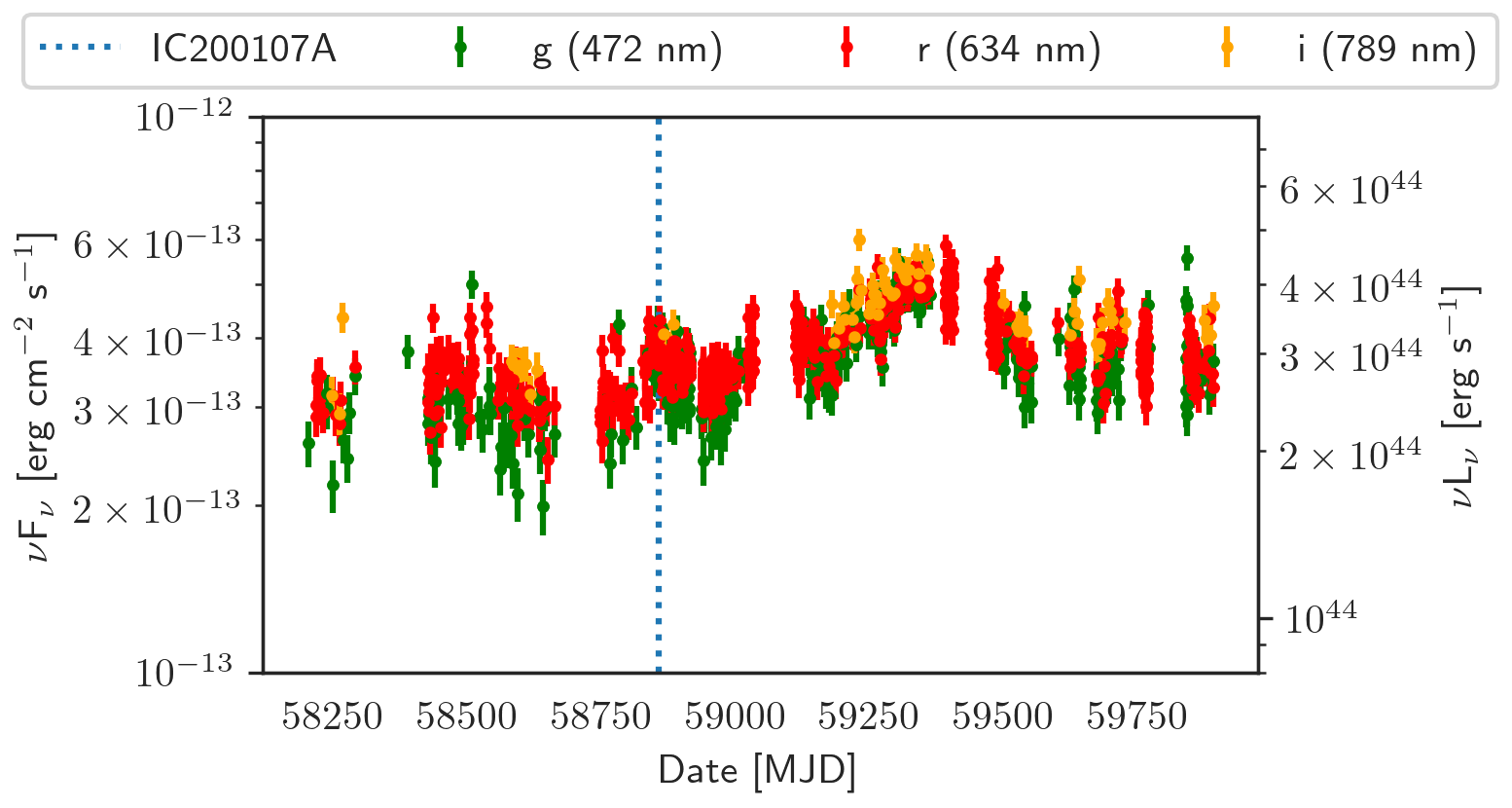}
	\caption{ZTF lightcurve of blazar BZB J0955+3551. The arrival time of neutrino IC200107A is marked with the vertical dashed line.}
	\label{fig:bzb_ligtcurve}
\end{figure}

More comprehensive multi-frequency modelling has confirmed that the detection of a neutrino alert from an extreme blazar is plausible, though the simultaneous X-ray flare may not be directly related to the neutrino production \citep{paliya_20, giommi_20, petropoulou_20}. The ZTF lightcurve for BZB J0955+3551 is shown in Figure \ref{fig:bzb_ligtcurve}. There is no evidence of any optical flaring on short or long timescales coincident with the detection of IC200107A. 

\section{Limits on Neutrino source populations}
\label{sec:limits}

With our program, we did not find any likely candidate counterparts from any population except TDEs.  We can consider limits that can be placed on other potential sources of astrophysical neutrinos given the non-detections. These limits will clearly not apply for TDEs, because for this population probable counterparts were detected. If AT 2019fdr is ultimately found not to be a TDE, for example if it instead an SLSN \citep{pitik_21}, the limits would not apply to that population either. It will however apply to all other populations which would be detected by ZTF, provided at least one detection occurred within the 14 day window after neutrino arrival time. Our search has no requirement that an optical lightcurve peaks after the neutrino detection, so these limits also apply to older/fading transients.

For each neutrino, we can consider the probability that an astrophysical counterpart would be detected. A counterpart could only be detected if a given IceCube neutrino was astrophysical, with this  as P$_{\textup{astro}}$ probability being reported by IceCube in GCN notices as the `\textit{signalness}' parameter. For each neutrino that was indeed astrophysical, the source could only then be detected if it lay within the area observed by ZTF. We can estimate this probability, P$_{\textup{obs}}$, by assuming that the 90\% probability is uniformly distributed across the rectangle reported by IceCube, A$_{\textup{IC}}$, such that:

\begin{equation}
P_{\textup{obs}} = 0.9 \times \frac{A_{\textup{ZTF}}}{A_{\textup{IC}}}
\label{eq:p_obs}
\end{equation}
where A$_{\textup{ZTF}}$ is the area observed by ZTF after accounting for detector chip gaps.

The probability to find an optical counterpart is then given by the joint probability that the neutrino is astrophysical, P$_{\textup{astro}}$, that the astrophysical source lay in the observed ZTF area, P$_{\textup{obs}}$, and the probability that a given counterpart would be detectable with our program $P_{\textup{detectable}}$. The values of $P_{\textup{astro}}$ and $P_{\textup{obs}}$ for each alert are given in Table \ref{tab:nu_alert_probs}. 

\begin{table}
	\centering
	\begin{tabular}{||c | c c | c ||} 
		\hline
		\textbf{Event} & P$_{\textup{astro}}$ & P$_{\textup{obs}}$ & $P_{\textup{astro}} \times P_{\textup{obs}}$ \\
		\hline
		IC190503A & 0.36 & 0.64 & 0.23 \\ 
		IC190619A & 0.55 & 0.71 & 0.39 \\ 
		IC190730A & 0.67 & 0.75 & 0.50 \\ 
		IC190922B & 0.51 & 0.82 & 0.42 \\ 
		IC191001A & 0.59 & 0.81 & 0.48 \\ 
		IC200107A & 0.50 & 0.74 & 0.37 \\ 
		IC200109A & 0.77 & 0.89 & 0.69 \\ 
		IC200117A & 0.38 & 0.84 & 0.32 \\ 
		IC200512A & 0.32 & 0.85 & 0.27 \\ 
		IC200530A & 0.59 & 0.78 & 0.46 \\ 
		IC200620A & 0.32 & 0.65 & 0.21 \\ 
		IC200916A & 0.32 & 0.77 & 0.25 \\ 
		IC200926A & 0.44 & 0.66 & 0.29 \\ 
		IC200929A & 0.47 & 0.70 & 0.33 \\ 
		IC201007A & 0.88 & 0.87 & 0.77 \\ 
		IC201021A & 0.30 & 0.82 & 0.25 \\ 
		IC201130A & 0.15 & 0.75 & 0.11 \\ 
		IC201209A & 0.19 & 0.61 & 0.12 \\ 
		IC201222A & 0.53 & 0.82 & 0.43 \\ 
		IC210210A & 0.65 & 0.67 & 0.43 \\ 
		IC210510A & 0.28 & 0.82 & 0.23 \\ 
		IC210629A & 0.35 & 0.69 & 0.24 \\ 
		IC210811A & 0.66 & 0.76 & 0.50 \\ 
		IC210922A & 0.93 & 0.67 & 0.62 \\ 
		
		\hline
	\end{tabular}
	\caption{Probability of finding a counterpart for each neutrino, assuming counterparts are sufficiently bright to be detected by our ZTF neutrino follow-up program.}
	\label{tab:nu_alert_probs}
\end{table}

The detectable probability will depend on the selection efficiency, $\epsilon_{\textup{det}}$, of our program. This selection efficiency in turn depends on the apparent magnitude of the electromagnetic counterpart. Motivated by our classification efficiency in Figure \ref{fig:completeness}, we assume completeness for objects brighter than 19.5 mag, and a classification efficiency of 68\% for objects fainter than this (this assumption is illustrated with the red dashed line in Figure \ref{fig:completeness}). We additionally assume a conservative 95\% detection efficiency for sources to be found by our pipeline, if said source was imaged by the camera. Chip gaps in the detector are already accounted for in Equation \ref{eq:p_obs}. Because the detection efficiency will decrease as the objects approach the ZTF limiting magnitude of 21.5 for 300s exposures, we neglect objects fainter than 21 mag in our calculation:

\begin{equation}
	\epsilon_{\textup{selection}}(m) =  0.95 \times \left\{
	\begin{array}{ll}
		1.00 & m \leq 19.5 \\
		0.68 & 19.5 \leq m\leq 21.0\\
		0.00 & 21.0 \leq m \\
	\end{array} 
	\right.
\end{equation}

The fraction of astrophysical neutrino sources that are detected by our program will depend on the properties of a given population.  For a power law neutrino spectrum, the differential neutrino particle flux at Earth for a transient population with a redshift-independent luminosity distribution is proportional to:

\begin{equation}
	\frac{dF(z)}{dz} = \frac{dN(z)}{dEdAdtdz}\propto  \left[ \ (1+z)^{2 - \gamma} \times \frac{R(z)}{4 \pi D_{L}^{2}} \right] \frac{dV_{C}}{dz}
	\label{eq:nu_flux_per_z}
\end{equation}
where $\gamma$ is the intrinsic neutrino spectral index such that $\frac{d N_{\nu}}{dE} \propto E ^{-\gamma}$, R(z) =$\rho(z)/\rho(0)$ is the normalised source redshift evolution for a population with rate $\rho(z)$, $D_{L}$ is the luminosity distance and $V_{C}$ is the comoving volume (see e.g \cite{murase_07}). By normalising Equation \ref{eq:nu_flux_per_z}, we can derive a probability density function (PDF) for the redshift of detected neutrinos:

\begin{equation}
	P_{\textup{dist}} (z) = \frac{dF(z)}{dz} \left/ \left( \int_{0}^{\infty} \frac{dF(z)}{dz} dz \right) \right.
\end{equation}

\begin{figure}
	\centering \includegraphics[width=0.48\textwidth]{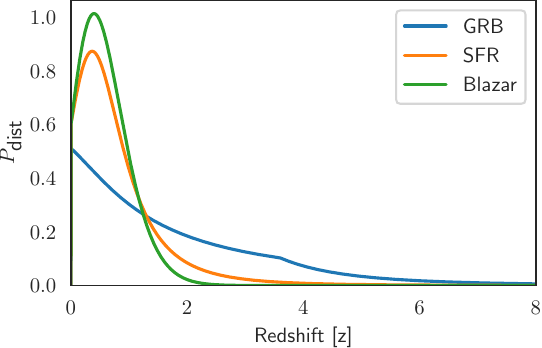}
	\caption{PDF for neutrino sources as a function of redshift, for both GRB-like and SFR-like source evolutions.}
	\label{fig:nu_cdf}
\end{figure}

PDFs for $P_{\textup{dist}} (z)$, calculated using the \emph{flarestack} code \citep{flarestack} for an E$^{-2}$ spectrum, are shown in Figure \ref{fig:nu_cdf} for redshift evolutions from a `GRB-like' population \citep{grb_lien_14} and from a Star-Formation-Rate population (`SFR-like') \citep{sfr_strolger_15}. We also show a blazar-like redshift evolution assuming a `primarily-density evolution' (PDE) following the formulation of \cite{ajello_15}, motivated by recent observations favouring such a distribution \citep{marcotulli_20}. The blazar redshift distribution is expected to vary with underlying source luminosity, so we assume a blazar gamma-ray luminosity of $L_{\gamma} \approx 10^{46}$ erg s$^{-1}$, motivated by observations of TXS 0506+056 \citep{padovani_18}, and the best fit parameters for a PDE  blazar evolution from \cite{ajello_15}.

It can be seen in Figure \ref{fig:nu_cdf} that GRB-like populations tend to be at greater distances than SFR-like ones, with GRB-like neutrinos being emitted from a median redshift of z = 1.34, whereas SFR-like neutrinos would have a median distance of z = 0.64 and blazar-like neutrinos would have a similar median of z = 0.57. This has a direct impact on the population properties compatible with our limits, because a neutrino population dominated by nearby sources will generally produce counterparts with brighter apparent magnitudes.

For a given source evolution, the probability of detecting a counterpart will then ultimately depend on the underlying luminosity function of the population. For an absolute magnitude, $M$, the counterpart detection probability is equal to the integrated product of the probability that a counterpart has a given redshift, $P_{\textup{dist}}(z)$, and the detection efficiency of our program for the apparent magnitude, $m(M, z)$, corresponding to that redshift:

\begin{equation}
	P_{\textup{detectable}} (M) = \int_{0}^{\infty} \left[ \epsilon_{\textup{det}}(m(M, z)) \times P_{\textup{dist}}(z) \right] dz
\end{equation}

\begin{figure}
	\centering \includegraphics[width=0.48\textwidth]{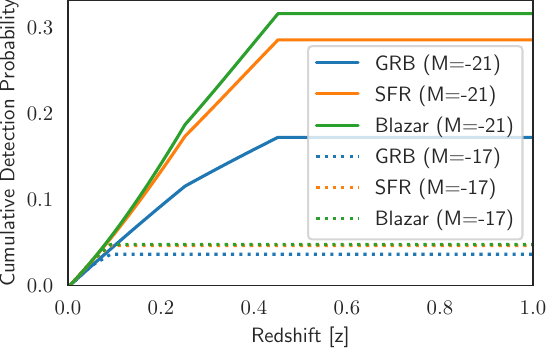}
	\caption{Cumulative counterpart detection probability as a function of redshift.}
	\label{fig:p_det}
\end{figure}

The impact of different evolutions and absolute magnitudes can be seen in Figure \ref{fig:p_det}. For sources with an absolute magnitude of $-21$, our program would be sensitive to counterparts up to a redshift of z $\approx$ 0.45, beyond which $m > 21$ so $\epsilon_{\textup{selection}} = 0$. For an SFR-like evolution, this would correspond to $P_{\textup{detectable}} (-21) = 26\%$, but for the higher-z GRB-like neutrino distribution, we would instead find $P_{\textup{detectable}} = 16\%$. For a fainter absolute magnitude of $-17$, our program would probe a much smaller volume up to redshift z $\approx$ 0.1, so then $P_{\textup{detectable}}$ would be 5\% and 4\% for SFR-like and GRB-like populations respectively.

Combining  these values, the joint probability for us to find a counterpart during a follow-up campaign is given by:

\begin{equation}
	P_{\textup{find}} (f, M) = P_{\textup{astro}} \times P_{\textup{obs}} \times P_{\textup{detectable}} (M) \times f (M)
\end{equation}
where $f$ is the fraction of astrophysical neutrino sources with an absolute magnitude equal to or brighter than $M$. The probability that no counterpart was detected in any of our 24 follow-up observations is then given by:

\begin{equation}
P_{\textup{no\_counterpart}} (M, f) = \prod_{i=1}^{24} \left( 1 -  P_{\textup{find, i} } (M, f)  \right)
\end{equation}

\begin{figure}
	\centering \includegraphics[width=0.45\textwidth]{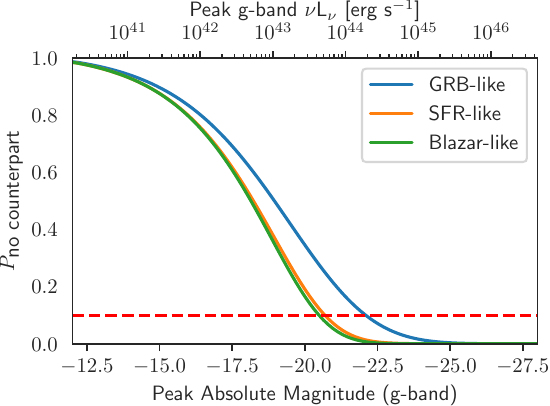}
	\caption{Probability of detecting no counterpart as a function of absolute magnitude for $f$=1. The dotted line corresponds to 90\% confidence.}
	\label{fig:p_no_det}
\end{figure}

The probability of no counterpart detection is given in Figure \ref{fig:p_no_det} as a function of $M$. The results of our program strongly disfavour scenarios where all neutrino sources have bright absolute magnitudes. The horizontal dashed line in  Figure \ref{fig:p_no_det} represents a 10\% chance of non-detection, and thus a 90\% confidence limit. We can use this threshold to set a limit on the luminosity function of neutrino sources, by choosing the appropriate fraction $f$ such that $P_{\textup{no counterpart}} (M, f) > 0.1$ .

\begin{figure}
	\centering \includegraphics[width=0.45\textwidth]{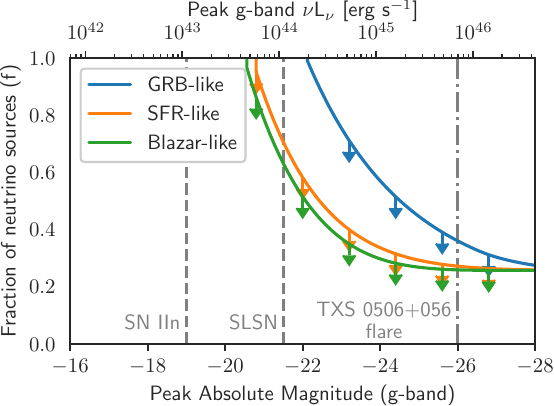}
	\caption{Upper limits (90\% CL) on the luminosity function of neutrino sources.}
	\label{fig:limit_abs_mag}
\end{figure}

These constraints on $f(M)$ at 90\% CL are illustrated in Figure \ref{fig:limit_abs_mag}, for the two source evolutions. These are generic constraints on the underlying luminosity function of neutrino sources, and are agnostic to the actual nature of the neutrino sources which follow the redshift evolutions. They constrain the aggregate neutrino flux emitted by e.g. a SFR-like population, and thus apply equally well to a composite neutrino flux with e.g. multiple SFR-like neutrino populations.  To the best knowledge of the authors, this is the first time generic constraints on the optical luminosity function of neutrino sources have been derived, though a similar procedure has already been used to derive limits from optical searches for counterparts to gravitational waves  \citep{kasliwal_20}. One novel consequence of these general limits are the first observational constraints on the contribution of the brightest superluminous supernova to the diffuse neutrino flux, under the assumption these trace the star formation rate. Objects brighter than an absolute magnitude of $-22$  can contribute no more than 58\% of the total astrophysical neutrino alerts if SFR-like. 

During the multi-wavelength flare of TXS 0506+056 which coincided with the detection of neutrino IC170922A, the source exhibited a g-band optical flux of  $\sim$6 mJy/m=14.5 \citep{ic_txs_mm_18}, more than double the mean archival g-band magnitude of 2.7 mJy/m=15.3 in Pan-STARRS1 \citep{panstarrs_16}. This would correspond to a transient flare of absolute magnitude M$_{g}\approx$-26 in difference imaging. Blazar flares such as this can contribute no more than 26\% of the total neutrino flux. 

It should be noted that none of these values account for the impact of dust extinction, which would shift the curves in Figure \ref{fig:limit_abs_mag} rightwards to higher luminosities. However, we do not expect that this would significantly impact the results presented here. Our limits are only valid for a given source evolutions, and would need to be adjusted for alternative ones. For example, if the neutrino source evolution were strongly negative (lower number density at higher redshift), a larger fraction of the flux would arise from local sources. Therefore, our limits would instead be more constraining.

It should also be noted that these limits assume that a given transient could pass our selection criteria outlined in Section \ref{sec:selection}, and therefore do not apply to extremely rapid transients such as GRB afterglows, which peak and fade on timescales $\lesssim$1 day. Such objects are not well captured by the ZTF public survey cadence or our typical neutrino follow-up observation cadence, and are unlikely to be detected multiple times in order to pass our selection criteria, so our detection efficiency will be somewhat lower.

\section{Conclusions}
\label{sec:conclusion}

The ZTF neutrino follow-up program coincided with the introduction of the upgraded IceCube alert selection, yielding one unretracted alert every 2 weeks and one ZTF  follow-up campaign every 4 weeks on average. The program resulted in the identification of two probable neutrino sources \citep{bran, tywin}, and the first limits on the optical luminosity function of neutrino sources.

\begin{figure}
	\centering \includegraphics[width=0.45\textwidth]{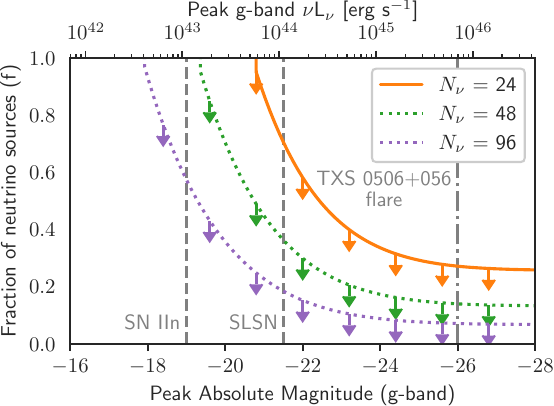}
	\caption{Upper limits (90\% CL) on the luminosity function of neutrino sources for an SFR-like evolution that would be derived for a ZTF neutrino sample that was twice ($N_{\nu}$=48) or four times ($N_{\nu}$=96) the size of the sample presented here.}
	\label{fig:limit_ztf_future}
\end{figure}

Though the limits presented here constrain only the very brightest transients such as superluminous supernovae (SLSNe), they will continue to become more stringent over time if no new counterparts are identified. As can be seen in Figure \ref{fig:limit_ztf_future}, extrapolating our analysis to a neutrino sample that was twice or four times as large would lead to substantially more constraining limits, and will be achieved on the present trajectory with 2 or 6 additional years of observations. 

Although the data analysis presented considered candidates detected up to 14 days after neutrino detection, our early real-time counterpart searches generally focussed on counterparts detected in the ToO observations scheduled for the first two nights after neutrino detection. Motivated by the systematic analysis performed here, and to improve sensitivity to time-delayed optical signatures such as neutrino emission from choked jets, we have modified our ToO observation strategy to better cover a range of transient timescales. We now trigger deep 300s in $g$ and $r$ band on the first night of observations to obtain deep upper limits or faint detections, and to additionally yield colour information for any active transient. However, we replaced our second pair of 300s exposures with a series of 30s exposures spread over subsequent nights, to complement the public survey and ensure good coverage of the photometric evolution of candidates. Forced photometry is only possible for images from the public survey after they have been published as part of the regular ZTF Data Releases, but with this ToO monitoring we can perform forced photometry analysis in real time \citep{ztffps}. We can also better prioritise spectroscopic follow-up with photometric classification.

One shortcoming of the ZTF program thus far has been the relatively poor sensitivity to very rapid transients such as GRB afterglows, owing to the median latency of 12.2 hours to first coverage. We plan to implement automated triggering with ZTF, similar to that operated by other observatories such as ASAS-SN \citep{necker_22}, enabling low-latency observations for at least some favourable neutrino alerts with appropriate accessibility. Dedicated analysis of low-latency follow-up campaigns would yield more stringent constraints on GRB afterglows as neutrino sources.

\begin{figure}
	\centering \includegraphics[width=0.45\textwidth]{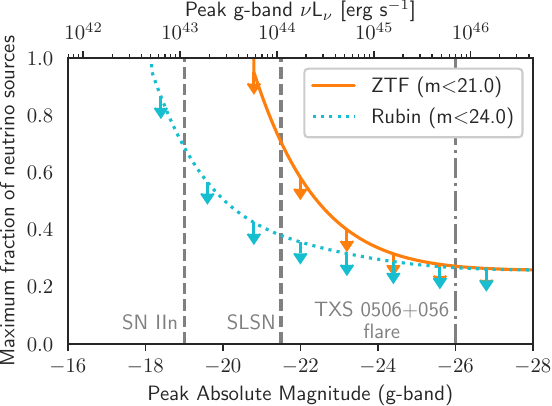}
	\caption{Upper limits (90\% CL) on the luminosity function for an SFR-like population with our sample of 24 observed neutrino alert and our classification efficiency (ZTF ), and limits that would be obtained for a comparable neutrino follow-up program with the upcoming Rubin Observatory.}
	\label{fig:limit_future}
\end{figure}

The results and analysis presented here can serve as a pathfinder for future triggered neutrino follow-up programs with wide-field instruments.  In particular, ToO observations with the upcoming Vera C. Rubin Observatory would offer an unprecedented opportunity to probe neutrino sources to much higher redshifts \citep{lsst_19}. Multi-band observation coverage would enable photometric classification of many candidates, substantially extending the classification efficiency presented in Figure \ref{fig:completeness} to much greater depths. An illustration of this is presented in Figure \ref{fig:limit_future}, assuming that the same neutrino sample in Table \ref{tab:nu_alerts} had instead been observed with the Rubin Observatory. For a comparable 60\% classification efficiency down to 24th mag, the corresponding limits on the neutrino luminosity function would be much more constraining for lower magnitudes. However, for very luminous optical sources such as blazar flares, the performance of both surveys for such a neutrino sample would be comparable. Given that there are only expected to be $\sim$12 astrophysical neutrinos in our sample, observations will never be able to overcome the 90\% limit from Poisson counting statistics even if they had a perfect 100\% efficiency. Instead, as seen in Figure \ref{fig:limit_ztf_future}, only larger neutrino samples can enable stricter limits on bright sources. 

Beyond optical observatories, similar electromagnetic neutrino follow-up programs are planned for telescopes currently under construction. These include near infra-red (NIR) wavelengths with WINTER \citep{winter_20, frostig_20}, at ultra-violet (UV) wavelengths  with ULTRASAT \citep{ultrasat_14}, and in gamma-rays with CTA \citep{cta_17, cta_21}. These new instruments, in concert  with the continuation of existing follow-up programs, will enable us to study the dynamic neutrino sky across the entire electromagnetic spectrum.

\section*{Acknowledgements}

R.S, and A.F. acknowledges support by the Initiative and Networking Fund of the Helmholtz Association through the Young Investigator Group program (A.F.). AF acknowledges funding from the German Science Foundation DFG, via the Collaborative Reasearch Center SFB1491 "Cosmic Interacting Matters - From Source to Signal. J.N. and S.R. acknowledges support by the Helmholtz Weizmann Research School on Multimessenger Astronomy, funded through the Initiative and Networking Fund of the Helmholtz Association, DESY, the Weizmann Institute, the Humboldt University of Berlin, and the University of Potsdam. ECK acknowledges support from the G.R.E.A.T research environment funded by {\em Vetenskapsr\aa det}, the Swedish Research Council, under project number 2016-06012, and support from The Wenner-Gren Foundations.

Based on observations obtained with the Samuel Oschin Telescope 48-inch and the 60-inch Telescope at the Palomar Observatory as part of the Zwicky Transient Facility project. ZTF is supported by the National Science Foundation under Grant No. AST-1440341 and AST-2034437, and a collaboration including Caltech, IPAC, the Weizmann Institute of Science, the Oskar Klein Center at Stockholm University, the University of Maryland, the University of Washington, Deutsches Elektronen-Synchrotron and Humboldt University, Los Alamos National Laboratories, the TANGO Consortium of Taiwan, the University of Wisconsin at Milwaukee, Lawrence Berkeley National Laboratories, Trinity College Dublin, Lawrence Livermore National Laboratories, IN2P3, France, the University of Warwick, the University of Bochum, and Northwestern University. Operations are conducted by COO, IPAC, and UW. SED Machine is based upon work supported by the National Science Foundation under Grant No. 1106171. The ZTF forced-photometry service was funded under the Heising-Simons Foundation grant \#12540303 (PI: Graham). 

The data presented herein were obtained in part at the W. M. Keck Observatory, which is operated as a scientific partnership among the California Institute of Technology, the University of California and the National Aeronautics and Space Administration. The Observatory was made possible by the generous financial support of the W. M. Keck Foundation. 

The authors wish to recognize and acknowledge the very significant cultural role and reverence that the summit of Maunakea has always had within the indigenous Hawaiian community.  We are most fortunate to have the opportunity to conduct observations from this mountain. 

Based on observations made with the Nordic Optical Telescope, owned in collaboration by the University of Turku and Aarhus University, and operated jointly by Aarhus University, the University of Turku and the University of Oslo, representing Denmark, Finland and Norway, the University of Iceland and Stockholm University at the Observatorio del Roque de los Muchachos, La Palma, Spain, of the Instituto de Astrofisica de Canarias. The data presented here were obtained in part with ALFOSC, which is provided by the Instituto de Astrofisica de Andalucia (IAA) under a joint agreement with the University of Copenhagen and NOT.

This research made use of Astropy, a community-developed core Python package for Astronomy \citep{2018AJ....156..123A, 2013A&A...558A..33A}. This research made use of Astroquery \citep{2019AJ....157...98G}, of the NASA/IPAC Extragalactic Database (NED) which is operated by the Jet Propulsion Laboratory, California Institute of Technology, under contract with the National Aeronautics and Space Administration. 

%%%%%%%%%%%%%%%%%%%%%%%%%%%%%%%%%%%%%%%%%%%%%%%%%%
\section*{Data Availability}

 The data presented here, and the Python analysis code used to generate the figures and key results, are available on Github:
 
  \url{https://github.com/robertdstein/nuztfpaper}
 
\noindent They are also available directly from the author upon reasonable request. 

% The data presented here, and the Python analysis code used to generate the figures and key results, can be found on Github at \url{https://github.com/robertdstein/nuztfpaper}.

%%%%%%%%%%%%%%%%%%%% REFERENCES %%%%%%%%%%%%%%%%%%

% The best way to enter references is to use BibTeX:

\bibliographystyle{mnras}
\bibliography{ztfnu} % if your bibtex file is called example.bib

% Alternatively you could enter them by hand, like this:
% This method is tedious and prone to error if you have lots of references
%\begin{thebibliography}{99}
%\bibitem[\protect\citeauthoryear{Author}{2012}]{Author2012}
%Author A.~N., 2013, Journal of Improbable Astronomy, 1, 1
%\bibitem[\protect\citeauthoryear{Others}{2013}]{Others2013}
%Others S., 2012, Journal of Interesting Stuff, 17, 198
%\end{thebibliography}

%%%%%%%%%%%%%%%%%%%%%%%%%%%%%%%%%%%%%%%%%%%%%%%%%%

%%%%%%%%%%%%%%%%% APPENDICES %%%%%%%%%%%%%%%%%%%%%

\clearpage
\appendix

\section{Not followed up}

\begin{figure}
	\centering \includegraphics[width=0.45\textwidth]{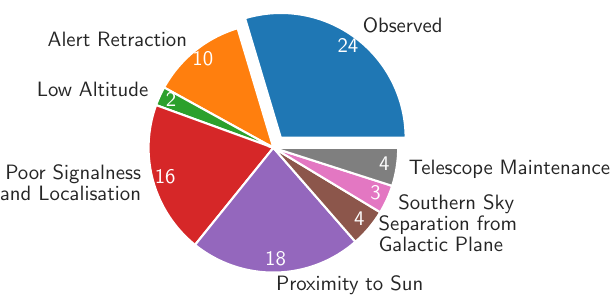}
	\caption{Breakdown of the neutrino follow-up program, as of 2021 Dec 31. }
	\label{fig:pie}
\end{figure}

Those alerts not observed by ZTF are summarised in Table \ref{tab:nu_non_observed}. Of those 57 alerts not followed up, the primary reasons were proximity to the Sun (18/55), alerts with poor localisation and low astrophysical probability (16/55) and alert retraction (10/55). The full breakdown of neutrino observations statistics can be seen in Figure \ref{fig:pie}.

\begin{table}
	\centering
	\begin{tabular}{||c c ||} 
		\hline
		\textbf{Cause} & \textbf{Events} \\
		\hline
		Alert Retraction & IC180423A \citep{ic180423a} \\ 
		& IC181031A \citep{ic181031a} \\ 
		& IC190205A \citep{ic190205a} \\ 
		& IC190529A \citep{ic190529a} \\ 
		& IC200120A \citep{ic200120a} \\ 
		& IC200728A \citep{ic200728a} \\ 
		& IC201115B \citep{ic201115b} \\ 
		& IC210213A \citep{ic210213a} \\ 
		& IC210322A \citep{ic210322a} \\ 
		& IC210519A \citep{ic210519a} \\ 
		\hline 
		Proximity to Sun & IC180908A \citep{ic180908a} \\ 
		& IC181014A \citep{ic181014a} \\ 
		& IC190124A \citep{ic190124a} \\ 
		& IC190704A \citep{ic190704a} \\ 
		& IC190712A \citep{ic190712a} \\ 
		& IC190819A \citep{ic190819a} \\ 
		& IC191119A \citep{ic191119a} \\ 
		& IC200227A \citep{ic200227a} \\ 
		& IC200421A \citep{ic200421a} \\ 
		& IC200615A \citep{ic200615a} \\ 
		& IC200806A \citep{ic200806a} \\ 
		& IC200921A \citep{ic200921a} \\ 
		& IC200926B \citep{ic200926b} \\ 
		& IC201014A \citep{ic201014a} \\ 
		& IC201115A \citep{ic201115a} \\ 
		& IC201221A \citep{ic201221a} \\ 
		& IC211117A \citep{ic211117a} \\ 
		& IC211123A \citep{ic211123a} \\ 
		\hline 
		Low Altitude & IC191215A \citep{ic191215a} \\ 
		& IC211023A \citep{ic211023a} \\ 
		\hline 
		Southern Sky & IC190104A \citep{ic190104a} \\ 
		& IC190331A \citep{ic190331a} \\ 
		& IC190504A \citep{ic190504a} \\ 
		\hline 
		Separation from Galactic Plane & IC201114A \citep{ic201114a} \\ 
		& IC201120A \citep{ic201120a} \\ 
		& IC210516A \citep{ic210516a} \\ 
		& IC210730A \citep{ic210730a} \\ 
		\hline 
		Poor Signalness and Localisation & IC190221A \citep{ic190221a} \\ 
		& IC190629A \citep{ic190629a} \\ 
		& IC190922A \citep{ic190922a} \\ 
		& IC191122A \citep{ic191122a} \\ 
		& IC191204A \citep{ic191204a} \\ 
		& IC191231A \citep{ic191231a} \\ 
		& IC200410A \citep{ic200410a} \\ 
		& IC200425A \citep{ic200425a} \\ 
		& IC200523A \citep{ic200523a} \\ 
		& IC200614A \citep{ic200614a} \\ 
		& IC200911A \citep{ic200911a} \\ 
		& IC210503A \citep{ic210503a} \\ 
		& IC210608A \citep{ic210608a} \\ 
		& IC210717A \citep{ic210717a} \\ 
		& IC211125A \citep{ic211125a} \\ 
		& IC211216A \citep{ic211216a} \\ 
		\hline 
		Telescope Maintenance & IC181023A \citep{ic181023a} \\ 
		& IC211116A \citep{ic211116a} \\ 
		& IC211208A \citep{ic211208a} \\ 
		& IC211216B \citep{ic211216b} \\ 
		\hline 
		
	\end{tabular}
	\caption{Summary of the 57 neutrino alerts that were not followed up by ZTF since survey start on 2018 March 20.}
	\label{tab:nu_non_observed}
\end{table}

%% Candidate tables
%%%%%%%%%%%%%%%%%%%%%%%%%%%%%%%%%%%%%%%%%%%%%%%%%%
\clearpage
\section{Candidates}

Candidates from each neutrino follow-up program are listed in tables \ref{tab:ic190503a}-\ref{tab:ic210811a}. Those candidates mentioned in the main text are highlighted in bold. For four neutrino campaigns (IC200107A, IC201007A, IC201222A and IC210922A), no candidates were identified, and there are no corresponding lists.

\begin{table}
	\centering
	\begin{tabular}{||c | c c c ||} 
		\hline
		\textbf{ZTF Name} & \textbf{IAU Name} & \textbf{Classification} & \textbf{Peak Magnitude} \\
		\hline
		ZTF19aatqcwq & -- & AGN Variability & 20.6 (g) \\ 
		ZTF19aatqlwq & -- & AGN Variability & 21.2 (r) \\ 
		
	\end{tabular}
	\caption{Candidates for IC190503A.}
	\label{tab:ic190503a}
\end{table}

\begin{table}
	\centering
	\begin{tabular}{||c | c c c ||} 
		\hline
		\textbf{ZTF Name} & \textbf{IAU Name} & \textbf{Classification} & \textbf{Peak Magnitude} \\
		\hline
		ZTF18abolwbb & -- & AGN Variability & 19.4 (r) \\ 
		ZTF18abueqkl & AT 2020kqj & AGN Variability & 19.3 (g) \\ 
		ZTF18acehkni & -- & AGN Variability & 19.4 (r) \\ 
		ZTF18actxchc & -- & AGN Variability & 18.0 (g) \\ 
		ZTF19aadaszg & SN 2019rg & SN Ia & 15.9 (r) \\ 
		ZTF19aawnawu & -- & AGN Variability & 20.0 (g) \\ 
		ZTF19aaycone & -- & AGN Variability & 17.9 (g) \\ 
		ZTF19aaycool & -- & AGN Variability & 20.3 (g) \\ 
		ZTF19aaycosc & -- & AGN Variability & 19.3 (r) \\ 
		ZTF19aaycoxd & -- & AGN Variability & 20.3 (g) \\ 
		ZTF19abahiwr & AT 2019izf & Unclassified & 19.5 (r) \\ 
		ZTF19abahiya & -- & Unclassified & 19.6 (r) \\ 
		ZTF19abahizn & -- & AGN Variability & 19.7 (g) \\ 
		ZTF19abahjcp & -- & AGN Variability & 20.2 (g) \\ 
		ZTF19abahlep & -- & Unclassified & 20.8 (r) \\ 
		ZTF19abahlka & -- & AGN Variability & 19.8 (i) \\ 
		ZTF19abajnby & -- & AGN Variability & 20.0 (r) \\ 
		
	\end{tabular}
	\caption{Candidates for IC190619A.}
	\label{tab:ic190619a}
\end{table}

\begin{table}
	\centering
	\begin{tabular}{||c | c c c ||} 
		\hline
		\textbf{ZTF Name} & \textbf{IAU Name} & \textbf{Classification} & \textbf{Peak Magnitude} \\
		\hline
		ZTF19aanlzzk & -- & Artefact & 13.8 (g) \\ 
		
	\end{tabular}
	\caption{Candidates for IC190730A.}
	\label{tab:ic190730a}
\end{table}

\begin{table}
	\centering
	\begin{tabular}{||c | c c c ||} 
		\hline
		\textbf{ZTF Name} & \textbf{IAU Name} & \textbf{Classification} & \textbf{Peak Magnitude} \\
		\hline
		ZTF18acekfly & AT 2019kkd & AGN Variability & 18.5 (r) \\ 
		ZTF19abcejyp & AT 2019kkp & AGN Variability & 19.3 (r) \\ 
		\textbf{ZTF19abxtupj} & \textbf{SN 2019pqh} & \textbf{SN II/IIb} & \textbf{20.3 (r)} \\ 
		
	\end{tabular}
	\caption{Candidates for IC190922B.}
	\label{tab:ic190922b}
\end{table}

\begin{table}
	\centering
	\begin{tabular}{||c | c c c ||} 
		\hline
		\textbf{ZTF Name} & \textbf{IAU Name} & \textbf{Classification} & \textbf{Peak Magnitude} \\
		\hline
		ZTF18ablvxkp & -- & AGN Variability & 19.3 (r) \\ 
		ZTF18absoqfm & -- & AGN Variability & 19.0 (g) \\ 
		\textbf{ZTF19aapreis} & \textbf{AT 2019dsg} & \textbf{TDE} & \textbf{17.8 (g)} \\ 
		ZTF19abassjx & -- & AGN Variability & 19.4 (i) \\ 
		ZTF19abcdynm & -- & AGN Variability & 20.5 (g) \\ 
		ZTF19abexshr & -- & AGN Variability & 20.2 (r) \\ 
		ZTF19abjfikj & -- & AGN Variability & 20.9 (g) \\ 
		ZTF19abjflnc & -- & AGN Variability & 19.2 (i) \\ 
		ZTF19abjflrg & -- & AGN Variability & 21.3 (g) \\ 
		ZTF19abjfmem & -- & AGN Variability & 21.5 (g) \\ 
		ZTF19abwaurq & -- & Unclassified & 19.5 (r) \\ 
		ZTF19abzkexb & SN 2019qhl & SN Ia & 18.9 (g) \\ 
		ZTF19acbpqfn & AT 2019rsj & Unclassified & 20.4 (g) \\ 
		ZTF19acbpqui & -- & Unclassified & 20.5 (g) \\ 
		ZTF19acbwpqs & -- & AGN Variability & 19.9 (g) \\ 
		ZTF19acbxahc & -- & Unclassified & 21.1 (g) \\ 
		ZTF19acbxanz & -- & Unclassified & 20.6 (r) \\ 
		ZTF19acbxaqj & -- & Unclassified & 20.5 (r) \\ 
		ZTF19acbxauk & -- & Unclassified & 20.8 (g) \\ 
		ZTF19acbxbjq & AT 2019rsk & Unclassified & 20.3 (g) \\ 
		ZTF19accnqlc & -- & Unclassified & 20.2 (r) \\ 
		
	\end{tabular}
	\caption{Candidates for IC191001A.}
	\label{tab:ic191001a}
\end{table}

\begin{table}
	\centering
	\begin{tabular}{||c | c c c ||} 
		\hline
		\textbf{ZTF Name} & \textbf{IAU Name} & \textbf{Classification} & \textbf{Peak Magnitude} \\
		\hline
		ZTF18aaidhnq & -- & AGN Variability & 18.1 (r) \\ 
		ZTF18aceykyg & -- & AGN Variability & 19.0 (g) \\ 
		ZTF18adgvgdk & -- & AGN Variability & 19.3 (g) \\ 
		ZTF19aangwsm & -- & Artefact & 19.8 (g) \\ 
		ZTF19aapsgtb & -- & AGN Variability & 18.8 (r) \\ 
		ZTF19aarohku & -- & AGN Variability & 19.8 (r) \\ 
		ZTF19acmwlds & AT 2019yfm & Unclassified & 19.7 (g) \\ 
		ZTF19adcdxgc & -- & AGN Variability & 19.6 (g) \\ 
		ZTF20aaeunmm & -- & AGN Variability & 20.4 (g) \\ 
		ZTF20aaeuufe & AT 2019yii  & Unclassified & 20.4 (r) \\ 
		ZTF20aaevfrv & -- & Star & 20.7 (g) \\ 
		ZTF20aaevfth & AT 2020ux & Unclassified & 21.2 (g) \\ 
		ZTF20aaevfwa & AT 2019zxa & Unclassified & 20.6 (r) \\ 
		ZTF20aaevgvt & AT 2020uw & Artefact & 20.5 (r) \\ 
		ZTF20aagvvve & -- & Artefact & 19.7 (r) \\ 
		ZTF20aagvvvh & -- & Artefact & 19.8 (r) \\ 
		ZTF20aagvvvk & -- & Artefact & 19.9 (r) \\ 
		ZTF20aagvvvn & -- & Artefact & 20.0 (r) \\ 
		ZTF20aagwcup & AT 2020dtc & Artefact & 19.9 (r) \\ 
		ZTF20aagwcuq & -- & Unclassified & 20.0 (r) \\ 
		ZTF20aagwcuu & -- & Unclassified & 20.0 (r) \\ 
		ZTF20aagwcuv & -- & Unclassified & 19.9 (r) \\ 
		ZTF20aagxfta & -- & Unclassified & 19.9 (g) \\ 
		
	\end{tabular}
	\caption{Candidates for IC200109A.}
	\label{tab:ic200109a}
\end{table}

\begin{table}
	\centering
	\begin{tabular}{||c | c c c ||} 
		\hline
		\textbf{ZTF Name} & \textbf{IAU Name} & \textbf{Classification} & \textbf{Peak Magnitude} \\
		\hline
		ZTF19acxopgh & AT 2019zyu & Unclassified & 19.4 (r) \\ 
		ZTF19adceqeb & -- & AGN Variability & 19.6 (g) \\ 
		ZTF20aacztcp & AT 2020ko & AGN Variability & 19.0 (r) \\ 
		ZTF20aaglixd & AT 2020agt & Unclassified & 21.2 (g) \\ 
		
	\end{tabular}
	\caption{Candidates for IC200117A.}
	\label{tab:ic200117a}
\end{table}

\begin{table}
	\centering
	\begin{tabular}{||c | c c c ||} 
		\hline
		\textbf{ZTF Name} & \textbf{IAU Name} & \textbf{Classification} & \textbf{Peak Magnitude} \\
		\hline
		ZTF18aazvbyj & -- & Star & 17.5 (r) \\ 
		ZTF18abjnqos & -- & Star & 12.9 (r) \\ 
		ZTF18abmfxbh & -- & Artefact & 17.5 (r) \\ 
		ZTF18abmfzmm & -- & Artefact & 17.1 (r) \\ 
		ZTF19acgpzgi & -- & Artefact & 15.5 (g) \\ 
		ZTF20aazqsfe & -- & Star & 19.6 (g) \\ 
		
	\end{tabular}
	\caption{Candidates for IC200512A.}
	\label{tab:ic200512a}
\end{table}

\begin{table}
	\centering
	\begin{tabular}{||c | c c c ||} 
		\hline
		\textbf{ZTF Name} & \textbf{IAU Name} & \textbf{Classification} & \textbf{Peak Magnitude} \\
		\hline
		ZTF18aaimsgg & AT 2018lnq & Artefact & 16.6 (r) \\ 
		ZTF18aamjqes & AT 2020llg & AGN Variability & 16.9 (r) \\ 
		ZTF18aaneyxs & -- & Artefact & 14.6 (r) \\ 
		\textbf{ZTF18aavecmo} & \textbf{AT 2020llh} & \textbf{AGN Flare} & \textbf{19.6 (i)} \\ 
		ZTF18aazkjyd & -- & Artefact & 14.7 (r) \\ 
		\textbf{ZTF18abrwqpr} & \textbf{AT 2020lli} & \textbf{AGN Flare} & \textbf{19.6 (g)} \\ 
		ZTF19aaonfhr & AT 2020llj & AGN Variability & 20.4 (r) \\ 
		ZTF19aascfca & -- & AGN Variability & 20.7 (g) \\ 
		ZTF19aascffj & -- & AGN Variability & 20.0 (g) \\ 
		\textbf{ZTF19aatubsj} & \textbf{AT 2019fdr} & \textbf{TDE} & \textbf{17.9 (i)} \\ 
		ZTF19abregmj & AT 2020llk & AGN Variability & 19.9 (g) \\ 
		ZTF20aaifyfd & AT 2020lll & AGN Variability & 19.9 (g) \\ 
		ZTF20aaifyrs & SN 2020awa & SN Ia & 17.0 (r) \\ 
		ZTF20aarbktd & SN 2020djn & SN II & 18.0 (i) \\ 
		ZTF20aavnpug & AT 2020idu & Dwarf Nova & 15.9 (i) \\ 
		ZTF20aawyens & AT 2020lpp & AGN Variability & 19.7 (i) \\ 
		ZTF20aaxcdok & AT 2020lpq & Unclassified & 20.1 (r) \\ 
		ZTF20aaxyglx & AT 2020llm & AGN Variability & 20.3 (g) \\ 
		ZTF20abaofgz & AT 2020lpr & AGN Variability & 19.9 (r) \\ 
		\textbf{ZTF20abbpkpa} & \textbf{SN 2020lam} & \textbf{SN II} & \textbf{18.8 (g)} \\ 
		ZTF20abcnrcb & -- & AGN Variability & 19.3 (g) \\ 
		ZTF20abdnovz & -- & Star & 21.3 (r) \\ 
		ZTF20abdnowa & AT 2020lln & Artefact & 20.7 (g) \\ 
		ZTF20abdnowp & AT 2020llo & Unclassified & 21.1 (g) \\ 
		ZTF20abdnowx & -- & AGN Variability & 21.3 (g) \\ 
		ZTF20abdnoxe & -- & AGN Variability & 20.3 (g) \\ 
		ZTF20abdnoxm & AT 2020llp & Unclassified & 20.8 (g) \\ 
		ZTF20abdnoyu & AT 2020lps & Unclassified & 21.4 (g) \\ 
		ZTF20abdnozk & AT 2020llq & AGN Variability & 20.6 (r) \\ 
		ZTF20abdnpae & AT 2020lpt & Unclassified & 20.9 (g) \\ 
		ZTF20abdnpbp & AT 2020llr & AGN Variability & 20.7 (r) \\ 
		ZTF20abdnpbq & AT 2020lpw & AGN Variability & 21.0 (r) \\ 
		ZTF20abdnpbu & AT 2020lpx & Unclassified & 21.0 (g) \\ 
		\textbf{ZTF20abdnpdo} & \textbf{SN 2020lls} & \textbf{SN Ic} & \textbf{19.0 (r)} \\ 
		ZTF20abdqzjl & -- & Star & 20.4 (r) \\ 
		ZTF20abdqzjr & -- & AGN Variability & 21.1 (r) \\ 
		ZTF20abdqzkq & AT 2020lpu & Star & 20.7 (g) \\ 
		ZTF20abdqzkr & -- & AGN Variability & 21.1 (g) \\ 
		ZTF20abdrnjw & -- & Star & 21.3 (r) \\ 
		ZTF20abdrnlg & AT 2020lpv & Unclassified & 20.9 (r) \\ 
		ZTF20abdrnmp & -- & AGN Variability & 21.6 (r) \\ 
		
	\end{tabular}
	\caption{Candidates for IC200530A.}
	\label{tab:ic200530a}
\end{table}

\begin{table}
	\centering
	\begin{tabular}{||c | c c c ||} 
		\hline
		\textbf{ZTF Name} & \textbf{IAU Name} & \textbf{Classification} & \textbf{Peak Magnitude} \\
		\hline
		ZTF18acvhwtf & AT 2020ncs & AGN Variability & 19.7 (r) \\ 
		ZTF20abgvabi & AT 2020ncr & AGN Variability & 20.2 (r) \\ 
		
	\end{tabular}
	\caption{Candidates for IC200620A.}
	\label{tab:ic200620a}
\end{table}

\begin{table}
	\centering
	\begin{tabular}{||c | c c c ||} 
		\hline
		\textbf{ZTF Name} & \textbf{IAU Name} & \textbf{Classification} & \textbf{Peak Magnitude} \\
		\hline
		ZTF18acccxxf & AT 2020tnn & AGN Variability & 19.7 (g) \\ 
		ZTF18adbbnry & AT 2020tnn & AGN Variability & 19.8 (g) \\ 
		ZTF20acaapwk & SN 2020tno & SN Ia & 18.9 (r) \\ 
		ZTF20acaapwn & -- & Unclassified & 21.0 (g) \\ 
		ZTF20acaapwo & AT 2020tnp & Unclassified & 20.4 (r) \\ 
		ZTF20acayuno & -- & AGN Variability & 21.1 (r) \\ 
		
	\end{tabular}
	\caption{Candidates for IC200916A.}
	\label{tab:ic200916a}
\end{table}

\begin{table}
	\centering
	\begin{tabular}{||c | c c c ||} 
		\hline
		\textbf{ZTF Name} & \textbf{IAU Name} & \textbf{Classification} & \textbf{Peak Magnitude} \\
		\hline
		ZTF18achvmdz & -- & AGN Variability & 18.9 (i) \\ 
		ZTF18acwfrle & -- & Star & 15.4 (g) \\ 
		
	\end{tabular}
	\caption{Candidates for IC200926A.}
	\label{tab:ic200926a}
\end{table}

\begin{table}
	\centering
	\begin{tabular}{||c | c c c ||} 
		\hline
		\textbf{ZTF Name} & \textbf{IAU Name} & \textbf{Classification} & \textbf{Peak Magnitude} \\
		\hline
		\textbf{ZTF20aamoxyt} & -- & \textbf{AGN Flare} & \textbf{19.8 (g)} \\ 
		
	\end{tabular}
	\caption{Candidates for IC200929A.}
	\label{tab:ic200929a}
\end{table}

\begin{table}
	\centering
	\begin{tabular}{||c | c c c ||} 
		\hline
		\textbf{ZTF Name} & \textbf{IAU Name} & \textbf{Classification} & \textbf{Peak Magnitude} \\
		\hline
		ZTF18abmkdiy & AT 2019cvb & AGN Variability & 18.7 (i) \\ 
		ZTF20abfaado & AT 2020nbr & Star & 19.3 (i) \\ 
		ZTF20acinqzo & -- & AGN Variability & 19.6 (i) \\ 
		ZTF20acmxnpa & AT 2020ybb & Unclassified & 20.6 (g) \\ 
		
	\end{tabular}
	\caption{Candidates for IC201021A.}
	\label{tab:ic201021a}
\end{table}

\begin{table}
	\centering
	\begin{tabular}{||c | c c c ||} 
		\hline
		\textbf{ZTF Name} & \textbf{IAU Name} & \textbf{Classification} & \textbf{Peak Magnitude} \\
		\hline
		ZTF17aadmvpm & -- & Artefact & 16.1 (g) \\ 
		\textbf{ZTF18abxrpgu} & \textbf{AT 2021ury} & \textbf{AGN Flare} & \textbf{18.8 (r)} \\ 
		ZTF18achpvrl & -- & AGN Variability & 19.1 (r) \\ 
		ZTF19aaagxcv & -- & AGN Variability & 18.4 (g) \\ 
		ZTF20aceidvg & -- & AGN Variability & 19.7 (g) \\ 
		ZTF20acmnnwf & -- & AGN Variability & 19.9 (r) \\ 
		ZTF20acuqdeu & AT 2020aehs & Unclassified & 19.8 (g) \\ 
		ZTF20acxbkpz & -- & Unclassified & 20.5 (r) \\ 
		
	\end{tabular}
	\caption{Candidates for IC201130A.}
	\label{tab:ic201130a}
\end{table}

\begin{table}
	\centering
	\begin{tabular}{||c | c c c ||} 
		\hline
		\textbf{ZTF Name} & \textbf{IAU Name} & \textbf{Classification} & \textbf{Peak Magnitude} \\
		\hline
		ZTF18abwhosy & -- & AGN Variability & 19.3 (r) \\ 
		ZTF20abvxjup & -- & AGN Variability & 20.0 (g) \\ 
		ZTF20acycunv & SN 2020addp & SN IIP & 19.4 (r) \\ 
		
	\end{tabular}
	\caption{Candidates for IC201209A.}
	\label{tab:ic201209a}
\end{table}

\begin{table}
	\centering
	\begin{tabular}{||c | c c c ||} 
		\hline
		\textbf{ZTF Name} & \textbf{IAU Name} & \textbf{Classification} & \textbf{Peak Magnitude} \\
		\hline
		ZTF19aaapmca & -- & AGN Variability & 18.6 (r) \\ 
		ZTF19aailrrn & -- & AGN Variability & 20.0 (g) \\ 
		ZTF19aasfvho & -- & AGN Variability & 19.4 (g) \\ 
		\textbf{ZTF19aasfvqm} & -- & \textbf{AGN Flare} & \textbf{18.2 (r)} \\ 
		ZTF20aadynqa & -- & AGN Variability & 20.1 (g) \\ 
		ZTF20aajcpde & -- & AGN Variability & 19.5 (g) \\ 
		ZTF21aafmkun & -- & AGN Variability & 19.4 (r) \\ 
		ZTF21aajxjmv & -- & Star & 21.3 (r) \\ 
		ZTF21aajxjmy & -- & Star & 21.1 (g) \\ 
		ZTF21aajxjnb & -- & AGN Variability & 22.1 (g) \\ 
		ZTF21aajxjnc & -- & AGN Variability & 21.7 (g) \\ 
		ZTF21aajxjrn & -- & AGN Variability & 20.1 (r) \\ 
		ZTF21aajxjrv & AT 2021clu & Unclassified & 20.9 (r) \\ 
		ZTF21aajxjry & AT 2021clv & Unclassified & 21.5 (r) \\ 
		ZTF21aajxjsa & -- & AGN Variability & 21.7 (r) \\ 
		ZTF21aajxkls & -- & AGN Variability & 21.1 (g) \\ 
		ZTF21aakiqpj & -- & Star & 22.1 (g) \\ 
		
	\end{tabular}
	\caption{Candidates for IC210210A.}
	\label{tab:ic210210a}
\end{table}

\begin{table}
	\centering
	\begin{tabular}{||c | c c c ||} 
		\hline
		\textbf{ZTF Name} & \textbf{IAU Name} & \textbf{Classification} & \textbf{Peak Magnitude} \\
		\hline
		ZTF19aadzayi & -- & Star & 15.0 (r) \\ 
		ZTF19aawqcum & -- & AGN Variability & 19.1 (g) \\ 
		ZTF20abhfiyd & -- & Star & 19.6 (g) \\ 
		ZTF20acinvxv & -- & Unclassified & 21.0 (r) \\ 
		ZTF20acinwlt & -- & AGN Variability & 21.0 (r) \\ 
		ZTF21aaiuekm & -- & Star & 19.5 (g) \\ 
		
	\end{tabular}
	\caption{Candidates for IC210510A.}
	\label{tab:ic210510a}
\end{table}

\begin{table}
	\centering
	\begin{tabular}{||c | c c c ||} 
		\hline
		\textbf{ZTF Name} & \textbf{IAU Name} & \textbf{Classification} & \textbf{Peak Magnitude} \\
		\hline
		ZTF18abteipt & AT 2019gnu & AGN Variability & 17.1 (r) \\ 
		ZTF21abecljv & AT 2021osi & AGN Variability & 19.8 (i) \\ 
		ZTF21abllruf & -- & Artefact & 17.5 (i) \\ 
		
	\end{tabular}
	\caption{Candidates for IC210629A.}
	\label{tab:ic210629a}
\end{table}

\begin{table}
	\centering
	\begin{tabular}{||c | c c c ||} 
		\hline
		\textbf{ZTF Name} & \textbf{IAU Name} & \textbf{Classification} & \textbf{Peak Magnitude} \\
		\hline
		ZTF20abjezpo & -- & Star & 19.7 (r) \\ 
		ZTF21absmcwm & -- & AGN Variability & 20.8 (g) \\ 
		
	\end{tabular}
	\caption{Candidates for IC210811A.}
	\label{tab:ic210811a}
\end{table}

%%%%%%%%%%%%%%%%%%%%%%%%%%%%%%%%%%

% Don't change these lines
\bsp	% typesetting comment
\label{lastpage}
\end{document}